\documentclass[reprint,amsmath,amssymb,aps]{revtex4-2}
\usepackage{lmodern}
\usepackage[latin9]{inputenc}
\usepackage{color}
\usepackage{float}
\usepackage{amsmath}
\usepackage{graphicx}
\usepackage[unicode=true,pdfusetitle,
 bookmarks=true,bookmarksnumbered=false,bookmarksopen=false,
 breaklinks=false,pdfborder={0 0 1},backref=false,colorlinks=false]
 {hyperref}
\hypersetup{
 colorlinks=true,allcolors=blue}
\usepackage{xr}

\makeatletter
\usepackage{dcolumn}
\usepackage{bm}

\makeatother

\begin{document}
\title{Catalysis by Dark States in Vibropolaritonic Chemistry}
\author{Matthew Du, Joel Yuen-Zhou}
\email{joelyuen@ucsd.edu}

\address{Department of Chemistry and Biochemistry, University of California
San Diego, La Jolla, California 92093, United States}
\date{\today}
\begin{abstract}
Collective strong coupling between a disordered ensemble of $N$ localized
molecular vibrations and a resonant optical cavity mode gives rise
to 2 polariton and $N-1\gg2$ dark modes. Thus, experimental changes
in thermally-activated reaction kinetics due to polariton formation
appear entropically unlikely and remain a puzzle. Here we show that
the overlooked dark modes, while parked at the same energy as bare
molecular vibrations, are robustly delocalized across $\sim$2-3 molecules,
yielding enhanced channels of vibrational cooling, concomitantly catalyzing
a chemical reaction. As an illustration, we theoretically
show a $\approx$50\% increase in an electron transfer rate due to enhanced
product stabilization. The reported effects can arise when the homogeneous linewidths of 
the dark modes are smaller than their energy spacings.
\end{abstract}
\maketitle
The past decade has seen much interest in the control of chemical
phenomena via the strong coupling of matter to confined electromagnetic
modes \citep{Agranovich2011rev,Coles2014,Ebbesen2016rev,Kowalewski2016,Chikkaraddy2016,Feist2017rev,Barnes2018,Vendrell2018prl,Herrera2018rev,Baranov2018rev,Ribeiro2018rev,Flick2018rev,Hertzog2019rev,Herrera2020rev,Coccia2020,Hoffmann2020,Antoniou2020,Haugland2020}.
An exciting prospect in this direction is vibropolaritonic chemistry,
that is, the use of collective vibrational strong coupling (VSC) \citep{Shalabney2015,Long2015,Casey2016}
to modify thermally-activated chemical reactivity without external
pumping (\textit{e.g.}, laser excitation) \citep{Hirai2020rev}. While
collective VSC involves a large number of molecules
per photon mode, it has been observed to substantially alter the kinetics
of organic substitution \citep{Thomas2016,Thomas2019}, cycloaddition
\citep{Hirai2020}, hydrolysis \citep{Hiura2019ChemRxiv}, enzyme
catalysis \citep{Vergauwe2019,Lather2021}, and crystallization \citep{Hirai2020chemRxiv},
among other electronic ground-state chemical processes.

However, such modified reactivity under VSC is still not well understood.
Studies \citep{Galego2019,Campos-Gonzalez-Angulo2020,Li2020,Zhdanov2020}
show that the observed kinetics cannot be explained with transition
state theory (TST) \citep{Hanggi1990rev}, the most commonly used
framework to predict and interpret reaction rates. Breakdowns of TST,
including recrossing the activation barrier \citep{Li2021}, deviation
from thermal equilibrium \citep{Li2020,Du2021}, and quantum/nonadiabatic
phenomena \citep{Campos-Gonzalez-Angulo2019,Phuc2020,Li2020,Du2021,Fischer2021},
have also been considered. The aforementioned works regard all $N$
vibrations coupled to the cavity mode to be identical. Under this
assumption, VSC forms two polaritons and $N-1$ optically dark vibrational
modes, where the latter remain unchanged from the cavity-free system.
It follows that VSC-induced changes to thermally-activated reactivity
must arise from the polaritons. In fact, one study from our group
highlighted the molecular parameter space where polaritons dominate
the kinetics \citep{Campos-Gonzalez-Angulo2019} with respect to dark
modes; however, this hypothesis has been questioned as entropically
unlikely \citep{Vurgaftman2020}.

Disorder, despite its ubiquity in molecular systems, has often been
ignored when modeling molecules under strong light-matter coupling.
Only recently has it been shown that the strong coupling of disordered
chromophores to an optical cavity mode can produce dark states which
are delocalized on multiple molecules \citep{Botzung2020,Scholes2020prs}
(hereafter, referred as semilocalized). This semilocalization is predicted
to improve or even enable coherent energy transport \citep{Botzung2020,Chavez2021}.
Other findings hint at adding sample impurities to help strong coupling
modify local molecular properties \citep{Sidler2021}.

\begin{figure}
\includegraphics{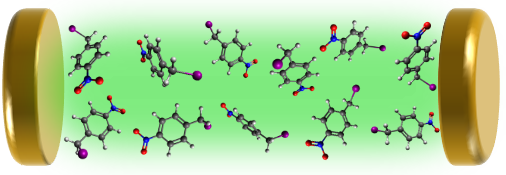}

\caption{\label{fig:setup} Schematic of the model setup. Inside an optical
cavity, a cavity mode collectively interacts with an energetically
disordered ensemble of molecular vibrations, each belonging to a separate
molecule. When a molecule reacts, its cavity-coupled vibrational mode
concomitantly experiences a displacement in equilibrium geometry.
Depicted here are molecules that undergo intramolecular electron transfer,
the reaction studied in this work.}
\end{figure}

In this Letter, we demonstrate that the VSC of a disordered molecular
ensemble (Fig. \ref{fig:setup}) can significantly modify the kinetics
of a thermally-activated chemical reaction. The altered reactivity
is attributed to the semilocalized dark modes. The semilocalization
affects the reaction rate by changing the efficiency with which a
reactive mode dissipates energy. 

Consider a disordered ensemble of $N$ molecular vibrations, respectively
corresponding to $N$ independent molecules, inside an optical cavity
(Fig. \ref{fig:setup}). The system is described by the Hamiltonian
$H=\hbar\omega_{c}a_{0}^{\dagger}a_{0}+\hbar\sum_{i=1}^{N}\omega_{i}a_{i}^{\dagger}a_{i}+\hbar\sum_{i=1}^{N}g_{i}(a_{i}^{\dagger}a_{0}+\text{h.c.})$.
Vibrational mode $i$ is represented by annihilation operator $a_{i}$
and has frequency $\omega_{i}=\overline{\omega}_{v}+\delta\omega_{i}$, where
$\overline{\omega}_{v}$ is the mean vibrational frequency and $\delta\omega_{i}$
is the frequency offset of mode $i$. Reflecting inhomogeneous broadening
(static diagonal disorder), $\delta\omega_{i}$ is a normally distributed
random variable with mean zero and standard deviation $\sigma_{v}$.
The cavity mode is represented by annihilation operator $a_{0}$,
has frequency $\omega_{c}$, and couples to vibration $i$ with strength
$g_{i}$. For simplicity, we hereafter take $g_{i}=g$ for all $i$.

Using $H$, we investigate the physicochemical properties of a disordered
molecular system under VSC. Unless otherwise noted, calculations assume
that the cavity is resonant with the average vibration, $\omega_{c}=\overline{\omega}_{v}$,
and couples to the vibrations with collective strength $g\sqrt{N}=8\sigma_{v}$ (for all $N$).
Numerical values reported below are obtained by averaging over 5000
disorder realizations, \emph{i.e.}, sets $\{\omega_{i}\}_{i=1}^{N}$.
In plots versus the $H$ eigenfrequencies, each data point is an average
over the $H$ eigenmodes---from all disorder realizations---whose
frequency lies in the bin $\left(\Delta\omega(l-\frac{1}{2}),\Delta\omega(l+\frac{1}{2})\right]$
for $\Delta\omega=10^{-1}\sigma_{v}$ and $l\in\mathbb{Z}$.

\begin{figure}
\includegraphics{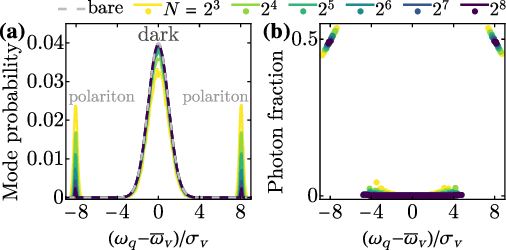}

\caption{\label{fig:prob-photonFrac}(a) Probability distribution and (b) photon
fraction of the eigenmodes of $H$. Values are plotted versus eigenfrequency
$\omega_{q}$ for various $N$. In (a), the gray dashed line is the
probability density function {[}$P(\omega_{q})=(\sigma_{v}\sqrt{2\pi})^{-1}\exp\left(-(\omega_{q}-\overline{\omega}_{v})^{2}/(2\sigma_{v}^{2})\right)$,
displayed in units of $1/\Delta\omega${]} of the bare vibrational
modes. }

\end{figure}

We first study the eigenmodes of $H$. Formally, mode $q=1,\dots,N+1$
is represented by operator $\alpha_{q}=\sum_{i=0}^{N}c_{qi}a_{i}$
and has frequency $\omega_{q}$. Figs. \ref{fig:prob-photonFrac}(a)
and \ref{fig:prob-photonFrac}(b) show the probability distribution
and photon fraction ($|c_{q0}|^{2}$), respectively, of the eigenmodes
with respect to eigenfrequency. The majority of modes form a broad
distribution in frequency around $\overline{\omega}_{v}$ and are
optically dark. A minority of modes are polaritons, which have frequency
$\overline{\omega}_{v}\pm8\sigma_{v}$, photon fraction $\approx0.5$,
and a lineshape minimally affected by inhomogeneous broadening \citep{Houdre1996}.
As $N$ rises, the eigenmodes become increasingly composed of dark
modes, whose probability distribution approaches that of the bare
vibrational modes {[}Fig. \ref{fig:prob-photonFrac}(a), gray dashed
line{]}. 

\begin{figure}
\includegraphics{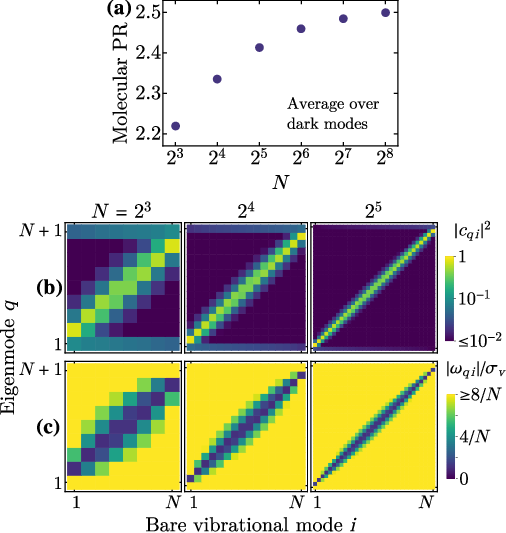}

\caption{\label{fig:molPR-arrays}(a) Average molecular PR of the dark modes
as a function of $N$. (b) Squared overlap $|c_{qi}|^{2}$ (log$_{10}$
scale) and (c) frequency difference $|\omega_{qi}|$ between each
eigenmode $q$ and each bare vibrational mode $i$. Each group of
modes is ordered from low to high frequency, \textit{i.e.}, dark (polariton)
modes have index $q=2,\dots,N$ ($q=1,N+1$). The quantities in (b)-(c)
are plotted for various $N$.}
\end{figure}

Next, we examine the delocalization of the dark modes. For the purpose
of studying chemical reactions, it is useful to compute the molecular
participation ratio (PR) \citep{Galego2017,Scholes2020prs}. This
measure, defined as
\begin{equation}
\text{molecular PR}=1/\sum_{i=1}^{N}\left|\frac{c_{qi}}{\sqrt{\sum_{i=1}^{N}|c_{qi}|^{2}}}\right|^{4}
\end{equation}
and analogous to the usual PR \citep{Kramer1993rev}, estimates the
number of molecules over which eigenmode $q$ is delocalized. According
to Fig. \ref{fig:molPR-arrays}(a), the average dark mode has molecular
PR $\sim$2-3, 
and this semilocalization persists as $N$ increases.
These phenomena were first noted independently by Scholes \citep{Scholes2020prs}
and Schachenmayer and co-workers \citep{Botzung2020}. For additional
insight, we plot the squared overlap {[}$|c_{qi}|^{2}$, Fig. \ref{fig:molPR-arrays}(b){]}
and frequency difference {[}$|\omega_{qi}|$, where $\omega_{qi}=\omega_{q}-\omega_{i}$;
Fig. \ref{fig:molPR-arrays}(c){]} between each dark mode and each
bare vibrational mode. The typical dark mode has sizable overlap with
the bare modes that are nearest to it in frequency. The frequency
difference between the dark mode and any of its major constituents
is $O(\sigma_{v}/N)$, \emph{i.e.}, negligible for large $N$. 

We now explore how VSC influences the kinetics of a thermally-activated
chemical reaction. Consider a reactive molecule under collective VSC
(Fig. \ref{fig:setup}). The molecule undergoes nonadiabatic intramolecular
electron transfer, and a cavity mode interacts collectively with a
reactive vibrational mode and $N-1$ nonreactive vibrational modes.
The model here considered is general enough that it should also be
applicable to the case of ``solvent-assisted VSC''
\citep{Lather2019,Schutz2020,Szidarovszky2020,Davidsson2020}.

To model the reaction of one molecule in the ensemble, we employ the
Hamiltonian $H_{\text{rxn}}=H+\sum_{X=R,P}|X\rangle\langle X|\{E_{X}+\hbar\omega_{r}[\lambda_{X}(a_{r}+a_{r}^{\dagger})+\lambda_{X}^{2}]\}+V+H_{s}^{(l)}$.
Note that vibrational and cavity modes are still described by $H$.
In writing $H_{\text{rxn}}$, we have changed the numerical index
of the bare vibrational mode ($i=1$) involved in the reaction to
the letter $r$ (\textit{i.e.}, $\omega_{1}\rightarrow\omega_{r}$,
$a_{1}\rightarrow a_{r}$); hereafter, we refer to this mode as $v_{r}$.
The electronic subspace consists of reactant $|R\rangle$ and product
$|P\rangle$ states. Electronic state $|X\rangle$ has energy $E_{X}$
and couples to $v_{r}$ with dimensionless strength $\lambda_{X}$.
As a result of the vibronic coupling, $v_{r}$ experiences a displacement
in its equilibrium position upon electron transfer. The interaction
between $|R\rangle$ and $|P\rangle$ is represented by $V=J_{RP}(|P\rangle\langle R|+\text{h.c.})$,
where $J_{RP}$ is the interaction strength. Through $H_{s}^{(l)}$
\citep{SM}, Hamiltonian $H_{\text{rxn}}$ also accounts for low-frequency
vibrational modes of the solvent that help mediate electron transfer. 
There is no direct coupling, though, of the cavity mode to 
the $|R\rangle \rightleftharpoons |P\rangle$ electronic transitions,
which we assume are dipole-forbidden.

Since we are considering a nonadiabatic reaction, we treat $V$ perturbatively
and calculate rates of reactive transitions between the zeroth-order
electronic-vibrational-cavity eigenstates of $H_{\text{rxn}}$. These
states take the form $|X,\chi\rangle=|X\rangle\otimes|\tilde{\chi}_{(X)}\rangle$.
Belonging to the subspace of vibrational and cavity modes, $|\tilde{\chi}_{(X)}\rangle=\left(\prod_{q=1}^{N+1}D_{q}^{\dagger}(\lambda_{Xq})\right)|\chi\rangle$
is a displaced Fock state with $m_{q}^{(\chi)}$ excitations in $H$
eigenmode $q$. The undisplaced Fock state $|\chi\rangle$ is an eigenstate
of $H$, and $D_{q}(\lambda)=\exp(\lambda\alpha_{q}^{\dagger}-\lambda^{*}\alpha_{q})$
is a displacement operator. Mode $q$ has equilibrium (dimensionless)
position $\lambda_{Xq}=\lambda_{X}c_{qr}(\omega_{r}/\omega_{q})$
when the system is in electronic state $|X\rangle$. Returning to
the electronic-vibrational-cavity state $|X,\chi\rangle$, we can
write its energy as $E_{(X,\chi)}=E_{X}+\sum_{q=1}^{N+1}m_{q}^{(\chi)}\hbar\omega_{q}+\Delta_{X}$,
where $\Delta_{X}=\lambda_{X}^{2}\hbar\omega_{r}-\hbar\sum_{q=1}^{N+1}|\lambda_{Xq}|^{2}\omega_{q}$
is the difference in reorganization energy---namely, that due to
vibronic coupling between $v_{r}$ and $|X\rangle$---with and without
VSC.

Following extensions \citep{Campos-Gonzalez-Angulo2019,Phuc2020,Du2021}
of Marcus-Levich-Jortner theory \citep{Marcus1964,Levich1966,Jortner1976}
to electron transfer under VSC, the rate of the reactive transition
from $|R,\chi\rangle$ to $|P,\chi'\rangle$ can be expressed as
\begin{align}
k_{(R,\chi)\rightarrow(P,\chi')} & =\mathcal{A}F_{\chi,\chi'}\exp\left(-\beta E_{a}^{\chi,\chi'}\right),\label{eq:k_ET}
\end{align}
where $\mathcal{A}=\sqrt{\pi\beta/\lambda_{s}}|J_{RP}|^{2}/\hbar$,
$\lambda_{s}$ is the reorganization energy associated with low-frequency
solvent modes \citep{SM}, and $\beta$ is the inverse temperature
For this transition, the activation energy is $E_{a}^{\chi,\chi'}=(E_{(P,\chi')}-E_{(R,\chi)}+\lambda_{s})^{2}/(4\lambda_{s})$.
Through the Franck-Condon (FC) factor $F_{\chi,\chi'}=|\langle\tilde{\chi}_{(R)}|\tilde{\chi}'_{(P)}\rangle|^{2}$,
the transition rate depends on the overlap between initial and final
vibrational-cavity states $|\tilde{\chi}_{(R)}\rangle$ and $|\tilde{\chi}'_{(P)}\rangle$,
respectively. It can be shown that the rate $k_{(P,\chi')\rightarrow(R,\chi)}$
corresponding to the backward transition, $|P,\chi'\rangle\rightarrow|R,\chi\rangle$,
is related to Eq. (\ref{eq:k_ET}) by detailed balance \citep{MayBook}.

We specifically study a reaction where, in the absence of light-matter
coupling, reactive transitions occur on the same timescale as internal
thermalization (\textit{i.e.}, thermalization of states having the
same electronic component). In such cases, internal thermal equilibrium
is not maintained throughout the reaction, and the reaction rate (\emph{i.e.},
net rate of reactant depletion) may not be approximated by a thermal
average of reactant-to-product transition rates. Instead, the reaction
rate can also depend on, \emph{e.g.}, backward reactive transitions
(from product to reactant) or vibrational relaxation. 

With this in mind, we numerically simulate the bare ($g=0$) and VSC
reactions using a kinetic model \citep{SM,Du2021}, which includes
forward and backward reactive transitions {[}Eq. (\ref{eq:k_ET}){]},
vibrational and cavity decay, and energy exchange among dark and polariton
states \citep{delPino2015,Xiang2018,Xiang2019jpca}. 
The third set of processes results from 
vibrational dephasing interactions (i.e., homogeneous broadening) 
of the molecular system \citep{delPino2015,SM}.
The reaction parameters \citep{SM}---in particular $\overline{\omega}_{v}=2000$
cm$^{-1}$, $\sigma_{v}=10$ cm$^{-1}$, $N\leq 32$, 
$g\sqrt{N}=8\sigma_{v}$ (for all $N$), $E_P-E_R=-0.6\overline{\omega}_{v}$, 
and the chosen temperature values---are such that the population
dynamics proceeds almost completely through states $|X,\chi\rangle$
with zero or one excitation in the vibrational-cavity modes. To reduce
computational cost, the kinetic model includes only these states,
which are denoted by $\chi=0$ and $\chi=1_{q}$, respectively, where
$q$ is an eigenmode of the vibrational-cavity subspace. The kinetic
master equation \citep{SM} is numerically solved with the initial
population being a thermal distribution of reactant states ($|R,\chi\rangle$).
Then, the \emph{apparent} reaction rate is obtained by fitting the
reactant population as a function of time \citep{SM}.
Note that, for large enough $N$ 
(we estimate $N > 72$ for the chosen parameters), 
the energy spacing between dark modes becomes smaller than their decay linewidths;
under this condition, our kinetic model is not valid \citep{SM}. 
Interestingly, such invalidation suggests that, within our model, 
VSC-modified chemistry might not occur in 
the weak light-matter coupling regime, 
where the polaritons and dark modes cannot be spectrally resolved.

\begin{figure}[t]
	\includegraphics{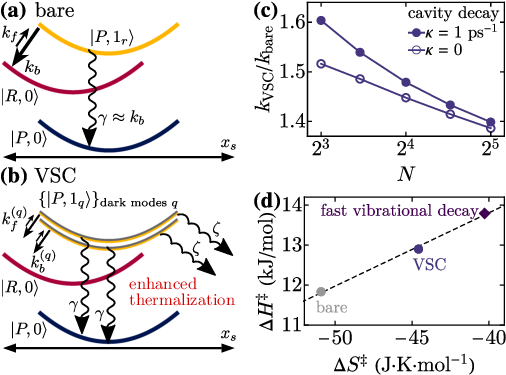}
	
	\caption{\label{fig:rxn}(a) Schematic of the bare reaction kinetics. Parabolas
		represent potential energy surfaces with respect to the effective
		low-frequency coordinate $x_{s}$. After the main reactive transition ($|R,0\rangle\rightarrow|P,1_{r}\rangle$)
		with rate $k_{f}$ (the transition $|R,0\rangle\rightarrow|P,0\rangle$ is not shown), 
		the product either reverts to the reactant ($|P,1_{r}\rangle\rightarrow|R,0\rangle$)
		at rate $k_{b}$ or vibrationally decays to its stable form ($|P,1_{r}\rangle\rightarrow|P,0\rangle$)
		at rate $\gamma\approx k_{b}$. (b) Schematic of the reaction kinetics
		under VSC. The half-yellow half-gray parabolas qualitatively represent
		potential energy surfaces of product states with one excitation in
		a dark mode ($|P,1_{q}\rangle$). The reaction proceeds via multiple
		reaction channels, each involving a reactive transition ($|R,0\rangle\rightarrow|P,1_{q}\rangle$)
		with rate $k_{f}^{(q)}<k_{f}$ to the product with one excitation
		in a dark mode. The total forward rate is approximately the bare rate
		$k_{f}$. In contrast, the vibrationally hot product formed from each
		reaction channel either returns to the reactant ($|P,1_{q}\rangle\rightarrow|R,0\rangle$)
		at rate $k_{b}^{(q)}<k_{b}$ or cools ($|P,1_{q}\rangle\rightarrow|P,0\rangle$)
		at the \emph{bare} molecular rate $\gamma$. There is also scattering
		(at effective rate $\zeta$) from dark modes with $c_{qr}\protect\not\approx0$
		to those with $c_{qr}\approx0$. Overall, VSC accelerates product
		thermalization, suppressing backward reactive transitions, and thus
		enhancing the net reaction rate. (c) $k_{\text{VSC}}/k_{\text{bare}}$ as a function of $N$ 
	    for fixed $g\sqrt{N}$ and various cavity decay rates $\kappa$. 
	    (d) Activation enthalpy $\Delta H^{\ddagger}$ versus activation entropy $\Delta S^{\ddagger}$
		for reactions with $N=32$: bare (gray circle), 
		VSC (purple circle, $\kappa=1$ ps$^{-1}$),
		and bare with vibrational decay rate $\gamma$ made 100 times faster
		(purple diamond). The black dashed line is a fit to the points shown.
		In (c) and (d), the individual rates ($k_{\text{VSC}}$, $k_{\text{bare}}$)
		and thermodynamic parameters ($\Delta H^{\ddagger}$, $\Delta S^{\ddagger}$)
		are averages over 5000 disorder realizations.}
\end{figure}

Even though we run the full numerical simulations as explained above,
we now introduce approximate models that shed conceptual intuition
on the calculated kinetics. First, the bare reaction can be essentially
captured by Fig. \ref{fig:rxn}(a), described as follows. Starting
from its vibrational ground state, the reactant converts to product
mainly by a $0\rightarrow1$ vibronic transition, which excites the
reactive mode and has rate $k_{f}\equiv k_{(R,0)\rightarrow(P,1_{r})}$,
where 
\begin{align}
k_{f} & =\mathcal{A}F_{0,1_{r}}\exp\left(-\beta E_{a}\right),\label{eq:k_f}
\end{align}
$E_{a}\equiv E_{a}^{0,1_{r}}$. The vibrationally hot product either
reverts to the reactant at rate $k_{b}\equiv k_{(P,1_{r})\rightarrow(R,0)}\gg k_{f}$,
where 
\begin{align}
k_{b} & =k_{f}\exp\left[\beta(E_{P}+\hbar\omega_{r}-E_{R})\right],\label{eq:k_b}
\end{align}
or decays to its vibrational ground state at rate $\gamma\approx k_{b}$.
Once the product reaches its vibrational ground state, it effectively
stops reacting due to the high reverse activation energy. This kinetic
scheme leads to a bare reaction rate \citep{SM}
\begin{equation}
k_{\text{bare}}^{(\text{analytical})}=k_{f}\left(\frac{\gamma}{\gamma+k_{b}}\right).\label{eq:k_bare}
\end{equation}

Second, under VSC, the primary reaction pathway of the bare case is
split into multiple pathways, each involving the (de)excitation of
a dark or polariton eigenmode $q$. For the VSC reaction channels,
the forward and backward rates take the form
\begin{align}
k_{f}^{(q)} & =\mathcal{A}F_{0,1_{q}}\exp\left(-\beta E_{a}^{(q)}\right),\label{eq:k_f_VSC}\\
k_{b}^{(q)} & =k_{f}^{(q)}\exp\left\{ \beta\left[E_{P}+\hbar\omega_{q}+\Delta_{P}-(E_{R}+\Delta_{R})\right]\right\} ,\label{eq:k_b_VSC}
\end{align}
respectively, where $k_{f}^{(q)}\equiv k_{(R,0)\rightarrow(P,1_{q})}$,
$k_{b}^{(q)}\equiv k_{(P,1_{q})\rightarrow(R,0)}$, and $E_{a}^{(q)}\equiv E_{a}^{0,1_{q}}$.
Now, consider the following argument, which holds strictly for large
$N$. As $N$ increases, the average bare mode becomes localized on
dark modes that have essentially the same frequency as it {[}Figs.
\ref{fig:molPR-arrays}(b)-\ref{fig:molPR-arrays}(c){]}, and its
overlap with the polariton modes vanishes {[}\emph{i.e.}, $c_{qr}\propto\frac{1}{\sqrt{N}}\to0$
for $q=1,N+1$; see Fig. \ref{fig:molPR-arrays}(b){]}. These observations
suggest that $c_{qr}\not\approx0$ only for modes $q$ that are dark
and have frequency $\omega_{q}\approx\omega_{r}$. It is then straightforward
to show that $\left.E_{a}^{(q)}\right|_{c_{qr}\not\approx0}\approx E_{a}$,
$F_{0,1_{q}}\approx|c_{qr}|^{2}F_{0,1_{r}}$, and
\begin{align}
k_{f/b}^{(q)} & \approx|c_{qr}|^{2}k_{f/b}.\label{eq:k_f/b}
\end{align}
Thus, VSC leads to reaction channels that have lower rates of reactive
transitions, due to changes not in activation energies but in FC factors,
which are smaller as a result of the semilocalization of dark modes.
From Eq. (\ref{eq:k_f/b}), it is evident that the total forward rate
is approximately that of the bare reaction ($\sum_{q=1}^{N+1}k_{f}^{(q)}\approx k_{f}$,
since $\sum_{q=1}^{N+1}|c_{qr}|^{2}=1$). However, once a forward
reactive transition happens---and a dark mode is excited---the product
either returns to the reactant at a \textit{reduced rate} ($k_{b}^{(q)}<k_{b}$)
or, due to the almost fully vibrational nature of the dark modes,
vibrationally decays to its stable form ($|P,0\rangle$) at essentially
the same \textit{bare rate} ($\gamma$). In other words, VSC suppresses
reverse reactive transitions by promoting the cooling of the reactive
mode upon product formation. In analogy to Eq. (\ref{eq:k_bare}),
we determine an effective rate for the VSC reaction \citep{SM}:
\begin{align}
k_{\text{VSC}}^{(\text{analytical})} & =k_{f}\left\langle \frac{\gamma}{\gamma+|c_{qr}|^{2}k_{b}}\right\rangle _{\text{dark modes }q},\label{eq:k_VSC}
\end{align}
where $\langle\cdot\rangle_{\text{dark modes }q}$ is a weighted average
over all dark modes $q$, each with weight $|c_{qr}|^{2}$. Since
$|c_{qr}|^{2}<1$ for all $q$, then $k_{\text{VSC}}^{(\text{analytical)}}>k_{\text{bare}}^{(\text{analytical})}$.
We emphasize that the major contributions to the average in Eq. (\ref{eq:k_VSC})
come from dark modes which are closest in frequency to the bare reactive
mode (see above). Further enhancement of the VSC reaction, beyond
that given by $k_{\text{VSC}}^{(\text{analytical})}$, occurs via
dissipative scattering from these dark modes to those with $c_{qr}\approx0$.
Said differently, the product is protected from reversion to reactant
when dark modes with relatively more reactive character lose their
energy to those with relatively less. 
Importantly, this scattering requires dark modes to be delocalized.
The VSC reaction kinetics, as
described above, is summarized in Fig. \ref{fig:rxn}(b).

Fig. \ref{fig:rxn}(c) shows the ratio of VSC reaction rate to bare
reaction rate, as determined from numerical kinetic simulations. As
we have shown analytically, VSC significantly accelerates the reaction
compared to the bare case. 
For $8 \leq N \leq 32$
(and $g\sqrt{N}$ held constant), the rate enhancement
is roughly
50\%. 
Notably, the effect of cavity decay on the reaction is minor and diminishes with $N$
[Fig. \ref{fig:rxn}(c)].
This behavior supports that
the reaction proceeds mainly through the dark
modes. 
The present scenario is quite generic 
and contrasts with our previous model where extreme
geometric parameters are needed for polaritons to dominate the VSC
kinetics \citep{Campos-Gonzalez-Angulo2019}. 

\begin{figure}
\includegraphics{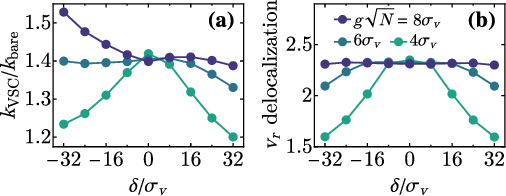}

\caption{\label{fig:ratio-deloc_g-detuning}(a) $k_{\text{VSC}}/k_{\text{bare}}$
and (b) $v_{r}$ delocalization, as a function of cavity detuning
$\delta$, for various collective light-matter coupling strengths
$g\sqrt{N}$ and fixed $N=32$. In (a), $k_{\text{VSC}}$ 
and $k_{\text{bare}}$ are averages
over 5000 disorder realizations.}
\end{figure}

We next look at
the
dependence on cavity detuning,
$\delta=\omega_{c}-\overline{\omega}_{v}$,
of the reaction rate and reactive-mode delocalization. 
The lattermost quantity is defined as $1/\sum_{q=1}^{N+1}|c_{qr}|^{4}$ (the PR of $v_{r}$ when the
mode is expressed in the eigenbasis of $H$).
We find, for various light-matter coupling strengths, that the reactive-mode
delocalization is maximum close to resonance and eventually decreases
with detuning [Fig. \ref{fig:ratio-deloc_g-detuning}(a)]. The rate enhancement due to VSC mostly follows the
same trend [Fig. \ref{fig:ratio-deloc_g-detuning}(b)]. 
Deviation from this trend 
at large negative detunings and collective light-matter couplings is attributed to 
polariton contributions to the rate which, 
as discussed in
\citep{Campos-Gonzalez-Angulo2019}, decrease as N increases \citep{SM}. 
The observed correlation between reactivity under VSC and 
delocalization of the reactive mode corroborates that 
the reaction is sped up by dark-mode semilocalization.
This mechanism is robust to moderate increases in inhomogeneous broadening
\citep{SM}.

For additional mechanistic insight into VSC catalysis and following
the procedures in \citep{Thomas2016,Lather2019,Hirai2020}, we plot
in Fig. \ref{fig:rxn}(d) the activation enthalpy ($\Delta H^{\ddagger}$)
versus activation entropy ($\Delta S^{\ddagger}$) for multiple cases
of VSC and bare reactions. The thermodynamic parameters of activation
are computed by calculating the apparent reaction rate for additional
temperatures and fitting the obtained values to the Eyring-Polanyi
equation \citep{SM}. This fit indicates that changes in effective
parameters $\Delta H^{\ddagger}$ and $\Delta S^{\ddagger}$ can result
from dynamical effects such as accelerated vibrational decay, rather
than from potential energy changes.

In conclusion, we show that, by forming semilocalized dark modes,
the VSC of a disordered molecular ensemble can modify the kinetics
of a thermally-activated chemical reaction. For a reactive molecule
under collective VSC, we find that the electron transfer rate is significantly
increased. The spreading of reactive character across dark modes,
as well as the dissipative scattering among these modes, allows the
reactive mode to thermalize more efficiently once the product is formed,
suppressing dynamical effects, such as reversion to reactant. 
Although experimental characterization of dark states remains a challenge, 
the phenomena proposed here might be verified using 
nonlinear infrared spectroscopy to measure populations \citep{Xiang2018,Ribeiro2018vp} 
and spatially resolved energy transport measurements to detect delocalization \citep{Botzung2020}.
The main mechanisms operating in our model
do not seem to be limited to nonadiabatic reactions and 
might have generalizations in adiabatic reactions; 
these will be explored in future work.
Given that these mechanisms only rely on collective VSC, 
they should also be operative in the cavity-free polaritonic architectures \citep{Canales2021}, 
although experiments along this front have so far not been reported.
More broadly, our work highlights that
the previously overlooked dark states are the entropically likely
channels through which collective light-matter interaction can control
chemistry.

\begin{acknowledgments}
\textcolor{black}{We are grateful to }Jorge Campos-Gonzalez-Angulo, 
Arghadip Koner, Luis Mart\'inez-Mart\'inez, Kai Schwennicke, Stephan
van den Wildenberg, and Garret Wiesehan for useful discussions. \textcolor{black}{This
work employed computational resources of the }Extreme Science and
Engineering Discovery Environment (XSEDE), which is supported by National
Science Foundation Grant No. ACI-1548562, under allocation No. TG-ASC150024.
We also thank Marty Kandes, Nicole Wolter, and Mahidhar Tatineni for assistance in using
these resources. 
Acknowledgment is made to the donors of The 
American Chemical Society Petroleum Research Fund for partial 
support of this research through the ACS PRF 60968-ND6 award. 
M.D. is also supported by a UCSD Roger Tsien Fellowship.
\end{acknowledgments}

\nocite{AgarwalBook,Agranovich2003,Litinskaya2004,Dunkelberger2016,
Grafton2021,
Duan2017,
Kuhn2003,Vohringer1995,Chang1993,
Xiang2019sa,
BreuerBook,
Frankcombe2001,
AtkinsBook9th,Eyring1935jcp,Evans1935}

\bibliographystyle{apsrev4-2}

\begin{thebibliography}{76}%
\makeatletter
\providecommand \@ifxundefined [1]{%
 \@ifx{#1\undefined}
}%
\providecommand \@ifnum [1]{%
 \ifnum #1\expandafter \@firstoftwo
 \else \expandafter \@secondoftwo
 \fi
}%
\providecommand \@ifx [1]{%
 \ifx #1\expandafter \@firstoftwo
 \else \expandafter \@secondoftwo
 \fi
}%
\providecommand \natexlab [1]{#1}%
\providecommand \enquote  [1]{``#1''}%
\providecommand \bibnamefont  [1]{#1}%
\providecommand \bibfnamefont [1]{#1}%
\providecommand \citenamefont [1]{#1}%
\providecommand \href@noop [0]{\@secondoftwo}%
\providecommand \href [0]{\begingroup \@sanitize@url \@href}%
\providecommand \@href[1]{\@@startlink{#1}\@@href}%
\providecommand \@@href[1]{\endgroup#1\@@endlink}%
\providecommand \@sanitize@url [0]{\catcode `\\12\catcode `\$12\catcode
  `\&12\catcode `\#12\catcode `\^12\catcode `\_12\catcode `\%12\relax}%
\providecommand \@@startlink[1]{}%
\providecommand \@@endlink[0]{}%
\providecommand \url  [0]{\begingroup\@sanitize@url \@url }%
\providecommand \@url [1]{\endgroup\@href {#1}{\urlprefix }}%
\providecommand \urlprefix  [0]{URL }%
\providecommand \Eprint [0]{\href }%
\providecommand \doibase [0]{https://doi.org/}%
\providecommand \selectlanguage [0]{\@gobble}%
\providecommand \bibinfo  [0]{\@secondoftwo}%
\providecommand \bibfield  [0]{\@secondoftwo}%
\providecommand \translation [1]{[#1]}%
\providecommand \BibitemOpen [0]{}%
\providecommand \bibitemStop [0]{}%
\providecommand \bibitemNoStop [0]{.\EOS\space}%
\providecommand \EOS [0]{\spacefactor3000\relax}%
\providecommand \BibitemShut  [1]{\csname bibitem#1\endcsname}%
\let\auto@bib@innerbib\@empty
\bibitem [{\citenamefont {Agranovich}\ \emph {et~al.}(2011)\citenamefont
  {Agranovich}, \citenamefont {Gartstein},\ and\ \citenamefont
  {Litinskaya}}]{Agranovich2011rev}%
  \BibitemOpen
  \bibfield  {author} {\bibinfo {author} {\bibfnamefont {V.~M.}\ \bibnamefont
  {Agranovich}}, \bibinfo {author} {\bibfnamefont {Y.~N.}\ \bibnamefont
  {Gartstein}},\ and\ \bibinfo {author} {\bibfnamefont {M.}~\bibnamefont
  {Litinskaya}},\ }\href {https://doi.org/10.1021/cr100156x} {\bibfield
  {journal} {\bibinfo  {journal} {Chem. Rev.}\ }\textbf {\bibinfo {volume}
  {111}},\ \bibinfo {pages} {5179} (\bibinfo {year} {2011})}\BibitemShut
  {NoStop}%
\bibitem [{\citenamefont {Coles}\ \emph {et~al.}(2014)\citenamefont {Coles},
  \citenamefont {Somaschi}, \citenamefont {Michetti}, \citenamefont {Clark},
  \citenamefont {Lagoudakis}, \citenamefont {Savvidis},\ and\ \citenamefont
  {Lidzey}}]{Coles2014}%
  \BibitemOpen
  \bibfield  {author} {\bibinfo {author} {\bibfnamefont {D.~M.}\ \bibnamefont
  {Coles}}, \bibinfo {author} {\bibfnamefont {N.}~\bibnamefont {Somaschi}},
  \bibinfo {author} {\bibfnamefont {P.}~\bibnamefont {Michetti}}, \bibinfo
  {author} {\bibfnamefont {C.}~\bibnamefont {Clark}}, \bibinfo {author}
  {\bibfnamefont {P.~G.}\ \bibnamefont {Lagoudakis}}, \bibinfo {author}
  {\bibfnamefont {P.~G.}\ \bibnamefont {Savvidis}},\ and\ \bibinfo {author}
  {\bibfnamefont {D.~G.}\ \bibnamefont {Lidzey}},\ }\href
  {https://doi.org/10.1038/nmat3950
  http://www.nature.com/nmat/journal/v13/n7/abs/nmat3950.html#supplementary-information}
  {\bibfield  {journal} {\bibinfo  {journal} {Nat. Mater.}\ }\textbf {\bibinfo
  {volume} {13}},\ \bibinfo {pages} {712} (\bibinfo {year} {2014})}\BibitemShut
  {NoStop}%
\bibitem [{\citenamefont {Ebbesen}(2016)}]{Ebbesen2016rev}%
  \BibitemOpen
  \bibfield  {author} {\bibinfo {author} {\bibfnamefont {T.~W.}\ \bibnamefont
  {Ebbesen}},\ }\href {https://doi.org/10.1021/acs.accounts.6b00295} {\bibfield
   {journal} {\bibinfo  {journal} {Acc. Chem. Res.}\ }\textbf {\bibinfo
  {volume} {49}},\ \bibinfo {pages} {2403} (\bibinfo {year}
  {2016})}\BibitemShut {NoStop}%
\bibitem [{\citenamefont {Kowalewski}\ \emph {et~al.}(2016)\citenamefont
  {Kowalewski}, \citenamefont {Bennett},\ and\ \citenamefont
  {Mukamel}}]{Kowalewski2016}%
  \BibitemOpen
  \bibfield  {author} {\bibinfo {author} {\bibfnamefont {M.}~\bibnamefont
  {Kowalewski}}, \bibinfo {author} {\bibfnamefont {K.}~\bibnamefont
  {Bennett}},\ and\ \bibinfo {author} {\bibfnamefont {S.}~\bibnamefont
  {Mukamel}},\ }\href {https://doi.org/10.1021/acs.jpclett.6b00864} {\bibfield
  {journal} {\bibinfo  {journal} {J. Phys. Chem. Lett.}\ }\textbf {\bibinfo
  {volume} {7}},\ \bibinfo {pages} {2050} (\bibinfo {year} {2016})}\BibitemShut
  {NoStop}%
\bibitem [{\citenamefont {Chikkaraddy}\ \emph {et~al.}(2016)\citenamefont
  {Chikkaraddy}, \citenamefont {de~Nijs}, \citenamefont {Benz}, \citenamefont
  {Barrow}, \citenamefont {Scherman}, \citenamefont {Rosta}, \citenamefont
  {Demetriadou}, \citenamefont {Fox}, \citenamefont {Hess},\ and\ \citenamefont
  {Baumberg}}]{Chikkaraddy2016}%
  \BibitemOpen
  \bibfield  {author} {\bibinfo {author} {\bibfnamefont {R.}~\bibnamefont
  {Chikkaraddy}}, \bibinfo {author} {\bibfnamefont {B.}~\bibnamefont
  {de~Nijs}}, \bibinfo {author} {\bibfnamefont {F.}~\bibnamefont {Benz}},
  \bibinfo {author} {\bibfnamefont {S.~J.}\ \bibnamefont {Barrow}}, \bibinfo
  {author} {\bibfnamefont {O.~A.}\ \bibnamefont {Scherman}}, \bibinfo {author}
  {\bibfnamefont {E.}~\bibnamefont {Rosta}}, \bibinfo {author} {\bibfnamefont
  {A.}~\bibnamefont {Demetriadou}}, \bibinfo {author} {\bibfnamefont
  {P.}~\bibnamefont {Fox}}, \bibinfo {author} {\bibfnamefont {O.}~\bibnamefont
  {Hess}},\ and\ \bibinfo {author} {\bibfnamefont {J.~J.}\ \bibnamefont
  {Baumberg}},\ }\href {https://doi.org/10.1038/nature17974} {\bibfield
  {journal} {\bibinfo  {journal} {Nature}\ }\textbf {\bibinfo {volume} {535}},\
  \bibinfo {pages} {127} (\bibinfo {year} {2016})}\BibitemShut {NoStop}%
\bibitem [{\citenamefont {Feist}\ \emph {et~al.}(2017)\citenamefont {Feist},
  \citenamefont {Galego},\ and\ \citenamefont {Garcia-Vidal}}]{Feist2017rev}%
  \BibitemOpen
  \bibfield  {author} {\bibinfo {author} {\bibfnamefont {J.}~\bibnamefont
  {Feist}}, \bibinfo {author} {\bibfnamefont {J.}~\bibnamefont {Galego}},\ and\
  \bibinfo {author} {\bibfnamefont {F.~J.}\ \bibnamefont {Garcia-Vidal}},\
  }\href {https://doi.org/10.1021/acsphotonics.7b00680} {\bibfield  {journal}
  {\bibinfo  {journal} {ACS Photonics}\ }\textbf {\bibinfo {volume} {5}},\
  \bibinfo {pages} {205} (\bibinfo {year} {2017})}\BibitemShut {NoStop}%
\bibitem [{\citenamefont {Barnes}\ \emph {et~al.}(2018)\citenamefont {Barnes},
  \citenamefont {Garc\'ia~Vidal},\ and\ \citenamefont {Aizpurua}}]{Barnes2018}%
  \BibitemOpen
  \bibfield  {author} {\bibinfo {author} {\bibfnamefont {B.}~\bibnamefont
  {Barnes}}, \bibinfo {author} {\bibfnamefont {F.}~\bibnamefont
  {Garc\'ia~Vidal}},\ and\ \bibinfo {author} {\bibfnamefont {J.}~\bibnamefont
  {Aizpurua}},\ }\href {https://doi.org/10.1021/acsphotonics.7b01609}
  {\bibfield  {journal} {\bibinfo  {journal} {ACS Photonics}\ }\textbf
  {\bibinfo {volume} {5}},\ \bibinfo {pages} {1} (\bibinfo {year}
  {2018})}\BibitemShut {NoStop}%
\bibitem [{\citenamefont {Vendrell}(2018)}]{Vendrell2018prl}%
  \BibitemOpen
  \bibfield  {author} {\bibinfo {author} {\bibfnamefont {O.}~\bibnamefont
  {Vendrell}},\ }\href {https://doi.org/10.1103/PhysRevLett.121.253001}
  {\bibfield  {journal} {\bibinfo  {journal} {Phys. Rev. Lett.}\ }\textbf
  {\bibinfo {volume} {121}},\ \bibinfo {pages} {253001} (\bibinfo {year}
  {2018})}\BibitemShut {NoStop}%
\bibitem [{\citenamefont {Herrera}\ and\ \citenamefont
  {Spano}(2018)}]{Herrera2018rev}%
  \BibitemOpen
  \bibfield  {author} {\bibinfo {author} {\bibfnamefont {F.}~\bibnamefont
  {Herrera}}\ and\ \bibinfo {author} {\bibfnamefont {F.~C.}\ \bibnamefont
  {Spano}},\ }\href {https://doi.org/10.1021/acsphotonics.7b00728} {\bibfield
  {journal} {\bibinfo  {journal} {ACS Photonics}\ }\textbf {\bibinfo {volume}
  {5}},\ \bibinfo {pages} {65} (\bibinfo {year} {2018})}\BibitemShut {NoStop}%
\bibitem [{\citenamefont {Baranov}\ \emph {et~al.}(2018)\citenamefont
  {Baranov}, \citenamefont {Wers\"all}, \citenamefont {Cuadra}, \citenamefont
  {Antosiewicz},\ and\ \citenamefont {Shegai}}]{Baranov2018rev}%
  \BibitemOpen
  \bibfield  {author} {\bibinfo {author} {\bibfnamefont {D.~G.}\ \bibnamefont
  {Baranov}}, \bibinfo {author} {\bibfnamefont {M.}~\bibnamefont {Wers\"all}},
  \bibinfo {author} {\bibfnamefont {J.}~\bibnamefont {Cuadra}}, \bibinfo
  {author} {\bibfnamefont {T.~J.}\ \bibnamefont {Antosiewicz}},\ and\ \bibinfo
  {author} {\bibfnamefont {T.}~\bibnamefont {Shegai}},\ }\href
  {https://doi.org/10.1021/acsphotonics.7b00674} {\bibfield  {journal}
  {\bibinfo  {journal} {ACS Photonics}\ }\textbf {\bibinfo {volume} {5}},\
  \bibinfo {pages} {24} (\bibinfo {year} {2018})}\BibitemShut {NoStop}%
\bibitem [{\citenamefont {Ribeiro}\ \emph
  {et~al.}(2018{\natexlab{a}})\citenamefont {Ribeiro}, \citenamefont
  {Mart\'inez-Mart\'inez}, \citenamefont {Du}, \citenamefont
  {Campos-Gonzalez-Angulo},\ and\ \citenamefont {Yuen-Zhou}}]{Ribeiro2018rev}%
  \BibitemOpen
  \bibfield  {author} {\bibinfo {author} {\bibfnamefont {R.~F.}\ \bibnamefont
  {Ribeiro}}, \bibinfo {author} {\bibfnamefont {L.~A.}\ \bibnamefont
  {Mart\'inez-Mart\'inez}}, \bibinfo {author} {\bibfnamefont {M.}~\bibnamefont
  {Du}}, \bibinfo {author} {\bibfnamefont {J.}~\bibnamefont
  {Campos-Gonzalez-Angulo}},\ and\ \bibinfo {author} {\bibfnamefont
  {J.}~\bibnamefont {Yuen-Zhou}},\ }\href {https://doi.org/10.1039/C8SC01043A}
  {\bibfield  {journal} {\bibinfo  {journal} {Chem. Sci.}\ }\textbf {\bibinfo
  {volume} {9}},\ \bibinfo {pages} {6325} (\bibinfo {year}
  {2018}{\natexlab{a}})}\BibitemShut {NoStop}%
\bibitem [{\citenamefont {Flick}\ \emph {et~al.}(2018)\citenamefont {Flick},
  \citenamefont {Rivera},\ and\ \citenamefont {Narang}}]{Flick2018rev}%
  \BibitemOpen
  \bibfield  {author} {\bibinfo {author} {\bibfnamefont {J.}~\bibnamefont
  {Flick}}, \bibinfo {author} {\bibfnamefont {N.}~\bibnamefont {Rivera}},\ and\
  \bibinfo {author} {\bibfnamefont {P.}~\bibnamefont {Narang}},\ }\href
  {https://doi.org/10.1515/nanoph-2018-0067} {\bibfield  {journal} {\bibinfo
  {journal} {Nanophotonics}\ }\textbf {\bibinfo {volume} {7}},\ \bibinfo
  {pages} {1479} (\bibinfo {year} {2018})}\BibitemShut {NoStop}%
\bibitem [{\citenamefont {Hertzog}\ \emph {et~al.}(2019)\citenamefont
  {Hertzog}, \citenamefont {Wang}, \citenamefont {Mony},\ and\ \citenamefont
  {B\"orjesson}}]{Hertzog2019rev}%
  \BibitemOpen
  \bibfield  {author} {\bibinfo {author} {\bibfnamefont {M.}~\bibnamefont
  {Hertzog}}, \bibinfo {author} {\bibfnamefont {M.}~\bibnamefont {Wang}},
  \bibinfo {author} {\bibfnamefont {J.}~\bibnamefont {Mony}},\ and\ \bibinfo
  {author} {\bibfnamefont {K.}~\bibnamefont {B\"orjesson}},\ }\href
  {https://doi.org/10.1039/C8CS00193F} {\bibfield  {journal} {\bibinfo
  {journal} {Chem. Soc. Rev.}\ }\textbf {\bibinfo {volume} {48}},\ \bibinfo
  {pages} {937} (\bibinfo {year} {2019})}\BibitemShut {NoStop}%
\bibitem [{\citenamefont {Herrera}\ and\ \citenamefont
  {Owrutsky}(2020)}]{Herrera2020rev}%
  \BibitemOpen
  \bibfield  {author} {\bibinfo {author} {\bibfnamefont {F.}~\bibnamefont
  {Herrera}}\ and\ \bibinfo {author} {\bibfnamefont {J.}~\bibnamefont
  {Owrutsky}},\ }\href {https://doi.org/10.1063/1.5136320} {\bibfield
  {journal} {\bibinfo  {journal} {J. Chem. Phys.}\ }\textbf {\bibinfo {volume}
  {152}},\ \bibinfo {pages} {100902} (\bibinfo {year} {2020})}\BibitemShut
  {NoStop}%
\bibitem [{\citenamefont {Coccia}\ \emph {et~al.}(2020)\citenamefont {Coccia},
  \citenamefont {Fregoni}, \citenamefont {Guido}, \citenamefont {Marsili},
  \citenamefont {Pipolo},\ and\ \citenamefont {Corni}}]{Coccia2020}%
  \BibitemOpen
  \bibfield  {author} {\bibinfo {author} {\bibfnamefont {E.}~\bibnamefont
  {Coccia}}, \bibinfo {author} {\bibfnamefont {J.}~\bibnamefont {Fregoni}},
  \bibinfo {author} {\bibfnamefont {C.~A.}\ \bibnamefont {Guido}}, \bibinfo
  {author} {\bibfnamefont {M.}~\bibnamefont {Marsili}}, \bibinfo {author}
  {\bibfnamefont {S.}~\bibnamefont {Pipolo}},\ and\ \bibinfo {author}
  {\bibfnamefont {S.}~\bibnamefont {Corni}},\ }\href
  {https://doi.org/10.1063/5.0027935} {\bibfield  {journal} {\bibinfo
  {journal} {J. Chem. Phys.}\ }\textbf {\bibinfo {volume} {153}},\ \bibinfo
  {pages} {200901} (\bibinfo {year} {2020})}\BibitemShut {NoStop}%
\bibitem [{\citenamefont {Hoffmann}\ \emph {et~al.}(2020)\citenamefont
  {Hoffmann}, \citenamefont {Lacombe}, \citenamefont {Rubio},\ and\
  \citenamefont {Maitra}}]{Hoffmann2020}%
  \BibitemOpen
  \bibfield  {author} {\bibinfo {author} {\bibfnamefont {N.~M.}\ \bibnamefont
  {Hoffmann}}, \bibinfo {author} {\bibfnamefont {L.}~\bibnamefont {Lacombe}},
  \bibinfo {author} {\bibfnamefont {A.}~\bibnamefont {Rubio}},\ and\ \bibinfo
  {author} {\bibfnamefont {N.~T.}\ \bibnamefont {Maitra}},\ }\href
  {https://doi.org/10.1063/5.0012723} {\bibfield  {journal} {\bibinfo
  {journal} {J. Chem. Phys.}\ }\textbf {\bibinfo {volume} {153}},\ \bibinfo
  {pages} {104103} (\bibinfo {year} {2020})}\BibitemShut {NoStop}%
\bibitem [{\citenamefont {Antoniou}\ \emph {et~al.}(2020)\citenamefont
  {Antoniou}, \citenamefont {Suchanek}, \citenamefont {Varner},\ and\
  \citenamefont {Foley}}]{Antoniou2020}%
  \BibitemOpen
  \bibfield  {author} {\bibinfo {author} {\bibfnamefont {P.}~\bibnamefont
  {Antoniou}}, \bibinfo {author} {\bibfnamefont {F.}~\bibnamefont {Suchanek}},
  \bibinfo {author} {\bibfnamefont {J.~F.}\ \bibnamefont {Varner}},\ and\
  \bibinfo {author} {\bibfnamefont {J.~J.}\ \bibnamefont {Foley}},\ }\href
  {https://doi.org/10.1021/acs.jpclett.0c02406} {\bibfield  {journal} {\bibinfo
   {journal} {J. Phys. Chem. Lett.}\ }\textbf {\bibinfo {volume} {11}},\
  \bibinfo {pages} {9063} (\bibinfo {year} {2020})}\BibitemShut {NoStop}%
\bibitem [{\citenamefont {Haugland}\ \emph {et~al.}(2020)\citenamefont
  {Haugland}, \citenamefont {Ronca}, \citenamefont {Kj{\o}nstad}, \citenamefont
  {Rubio},\ and\ \citenamefont {Koch}}]{Haugland2020}%
  \BibitemOpen
  \bibfield  {author} {\bibinfo {author} {\bibfnamefont {T.~S.}\ \bibnamefont
  {Haugland}}, \bibinfo {author} {\bibfnamefont {E.}~\bibnamefont {Ronca}},
  \bibinfo {author} {\bibfnamefont {E.~F.}\ \bibnamefont {Kj{\o}nstad}},
  \bibinfo {author} {\bibfnamefont {A.}~\bibnamefont {Rubio}},\ and\ \bibinfo
  {author} {\bibfnamefont {H.}~\bibnamefont {Koch}},\ }\href
  {https://doi.org/10.1103/PhysRevX.10.041043} {\bibfield  {journal} {\bibinfo
  {journal} {Phys. Rev. X}\ }\textbf {\bibinfo {volume} {10}},\ \bibinfo
  {pages} {041043} (\bibinfo {year} {2020})}\BibitemShut {NoStop}%
\bibitem [{\citenamefont {Shalabney}\ \emph {et~al.}(2015)\citenamefont
  {Shalabney}, \citenamefont {George}, \citenamefont {Hutchison}, \citenamefont
  {Pupillo}, \citenamefont {Genet},\ and\ \citenamefont
  {Ebbesen}}]{Shalabney2015}%
  \BibitemOpen
  \bibfield  {author} {\bibinfo {author} {\bibfnamefont {A.}~\bibnamefont
  {Shalabney}}, \bibinfo {author} {\bibfnamefont {J.}~\bibnamefont {George}},
  \bibinfo {author} {\bibfnamefont {J.}~\bibnamefont {Hutchison}}, \bibinfo
  {author} {\bibfnamefont {G.}~\bibnamefont {Pupillo}}, \bibinfo {author}
  {\bibfnamefont {C.}~\bibnamefont {Genet}},\ and\ \bibinfo {author}
  {\bibfnamefont {T.~W.}\ \bibnamefont {Ebbesen}},\ }\href
  {https://doi.org/10.1038/ncomms6981} {\bibfield  {journal} {\bibinfo
  {journal} {Nat. Commun.}\ }\textbf {\bibinfo {volume} {6}},\ \bibinfo {pages}
  {6} (\bibinfo {year} {2015})}\BibitemShut {NoStop}%
\bibitem [{\citenamefont {Long}\ and\ \citenamefont
  {Simpkins}(2015)}]{Long2015}%
  \BibitemOpen
  \bibfield  {author} {\bibinfo {author} {\bibfnamefont {J.~P.}\ \bibnamefont
  {Long}}\ and\ \bibinfo {author} {\bibfnamefont {B.~S.}\ \bibnamefont
  {Simpkins}},\ }\href {https://doi.org/10.1021/ph5003347} {\bibfield
  {journal} {\bibinfo  {journal} {ACS Photonics}\ }\textbf {\bibinfo {volume}
  {2}},\ \bibinfo {pages} {130} (\bibinfo {year} {2015})}\BibitemShut {NoStop}%
\bibitem [{\citenamefont {Casey}\ and\ \citenamefont
  {Sparks}(2016)}]{Casey2016}%
  \BibitemOpen
  \bibfield  {author} {\bibinfo {author} {\bibfnamefont {S.~R.}\ \bibnamefont
  {Casey}}\ and\ \bibinfo {author} {\bibfnamefont {J.~R.}\ \bibnamefont
  {Sparks}},\ }\href {https://doi.org/10.1021/acs.jpcc.6b10493} {\bibfield
  {journal} {\bibinfo  {journal} {J. Phys. Chem. C}\ }\textbf {\bibinfo
  {volume} {120}},\ \bibinfo {pages} {28138} (\bibinfo {year}
  {2016})}\BibitemShut {NoStop}%
\bibitem [{\citenamefont {Hirai}\ \emph
  {et~al.}(2020{\natexlab{a}})\citenamefont {Hirai}, \citenamefont
  {Hutchison},\ and\ \citenamefont {Uji-i}}]{Hirai2020rev}%
  \BibitemOpen
  \bibfield  {author} {\bibinfo {author} {\bibfnamefont {K.}~\bibnamefont
  {Hirai}}, \bibinfo {author} {\bibfnamefont {J.~A.}\ \bibnamefont
  {Hutchison}},\ and\ \bibinfo {author} {\bibfnamefont {H.}~\bibnamefont
  {Uji-i}},\ }\href {https://doi.org/10.1002/cplu.202000411} {\bibfield
  {journal} {\bibinfo  {journal} {ChemPlusChem}\ }\textbf {\bibinfo {volume}
  {85}},\ \bibinfo {pages} {1981} (\bibinfo {year}
  {2020}{\natexlab{a}})}\BibitemShut {NoStop}%
\bibitem [{\citenamefont {Thomas}\ \emph {et~al.}(2016)\citenamefont {Thomas},
  \citenamefont {George}, \citenamefont {Shalabney}, \citenamefont {Dryzhakov},
  \citenamefont {Varma}, \citenamefont {Moran}, \citenamefont {Chervy},
  \citenamefont {Zhong}, \citenamefont {Devaux}, \citenamefont {Genet},
  \citenamefont {Hutchison},\ and\ \citenamefont {Ebbesen}}]{Thomas2016}%
  \BibitemOpen
  \bibfield  {author} {\bibinfo {author} {\bibfnamefont {A.}~\bibnamefont
  {Thomas}}, \bibinfo {author} {\bibfnamefont {J.}~\bibnamefont {George}},
  \bibinfo {author} {\bibfnamefont {A.}~\bibnamefont {Shalabney}}, \bibinfo
  {author} {\bibfnamefont {M.}~\bibnamefont {Dryzhakov}}, \bibinfo {author}
  {\bibfnamefont {S.~J.}\ \bibnamefont {Varma}}, \bibinfo {author}
  {\bibfnamefont {J.}~\bibnamefont {Moran}}, \bibinfo {author} {\bibfnamefont
  {T.}~\bibnamefont {Chervy}}, \bibinfo {author} {\bibfnamefont
  {X.}~\bibnamefont {Zhong}}, \bibinfo {author} {\bibfnamefont
  {E.}~\bibnamefont {Devaux}}, \bibinfo {author} {\bibfnamefont
  {C.}~\bibnamefont {Genet}}, \bibinfo {author} {\bibfnamefont {J.~A.}\
  \bibnamefont {Hutchison}},\ and\ \bibinfo {author} {\bibfnamefont {T.~W.}\
  \bibnamefont {Ebbesen}},\ }\href {https://doi.org/10.1002/ange.201605504}
  {\bibfield  {journal} {\bibinfo  {journal} {Angew. Chem., Int. Ed.}\ }\textbf
  {\bibinfo {volume} {128}},\ \bibinfo {pages} {11634} (\bibinfo {year}
  {2016})}\BibitemShut {NoStop}%
\bibitem [{\citenamefont {Thomas}\ \emph {et~al.}(2019)\citenamefont {Thomas},
  \citenamefont {Lethuillier-Karl}, \citenamefont {Nagarajan}, \citenamefont
  {Vergauwe}, \citenamefont {George}, \citenamefont {Chervy}, \citenamefont
  {Shalabney}, \citenamefont {Devaux}, \citenamefont {Genet}, \citenamefont
  {Moran},\ and\ \citenamefont {Ebbesen}}]{Thomas2019}%
  \BibitemOpen
  \bibfield  {author} {\bibinfo {author} {\bibfnamefont {A.}~\bibnamefont
  {Thomas}}, \bibinfo {author} {\bibfnamefont {L.}~\bibnamefont
  {Lethuillier-Karl}}, \bibinfo {author} {\bibfnamefont {K.}~\bibnamefont
  {Nagarajan}}, \bibinfo {author} {\bibfnamefont {R.~M.~A.}\ \bibnamefont
  {Vergauwe}}, \bibinfo {author} {\bibfnamefont {J.}~\bibnamefont {George}},
  \bibinfo {author} {\bibfnamefont {T.}~\bibnamefont {Chervy}}, \bibinfo
  {author} {\bibfnamefont {A.}~\bibnamefont {Shalabney}}, \bibinfo {author}
  {\bibfnamefont {E.}~\bibnamefont {Devaux}}, \bibinfo {author} {\bibfnamefont
  {C.}~\bibnamefont {Genet}}, \bibinfo {author} {\bibfnamefont
  {J.}~\bibnamefont {Moran}},\ and\ \bibinfo {author} {\bibfnamefont {T.~W.}\
  \bibnamefont {Ebbesen}},\ }\href {https://doi.org/10.1126/science.aau7742}
  {\bibfield  {journal} {\bibinfo  {journal} {Science}\ }\textbf {\bibinfo
  {volume} {363}},\ \bibinfo {pages} {615} (\bibinfo {year}
  {2019})}\BibitemShut {NoStop}%
\bibitem [{\citenamefont {Hirai}\ \emph
  {et~al.}(2020{\natexlab{b}})\citenamefont {Hirai}, \citenamefont {Takeda},
  \citenamefont {Hutchison},\ and\ \citenamefont {Uji-i}}]{Hirai2020}%
  \BibitemOpen
  \bibfield  {author} {\bibinfo {author} {\bibfnamefont {K.}~\bibnamefont
  {Hirai}}, \bibinfo {author} {\bibfnamefont {R.}~\bibnamefont {Takeda}},
  \bibinfo {author} {\bibfnamefont {J.~A.}\ \bibnamefont {Hutchison}},\ and\
  \bibinfo {author} {\bibfnamefont {H.}~\bibnamefont {Uji-i}},\ }\href
  {https://doi.org/10.1002/ange.201915632} {\bibfield  {journal} {\bibinfo
  {journal} {Angew. Chem., Int. Ed.}\ }\textbf {\bibinfo {volume} {132}},\
  \bibinfo {pages} {5370} (\bibinfo {year} {2020}{\natexlab{b}})}\BibitemShut
  {NoStop}%
\bibitem [{\citenamefont {Hiura}\ \emph {et~al.}(2019)\citenamefont {Hiura},
  \citenamefont {Shalabney},\ and\ \citenamefont {George}}]{Hiura2019ChemRxiv}%
  \BibitemOpen
  \bibfield  {author} {\bibinfo {author} {\bibfnamefont {H.}~\bibnamefont
  {Hiura}}, \bibinfo {author} {\bibfnamefont {A.}~\bibnamefont {Shalabney}},\
  and\ \bibinfo {author} {\bibfnamefont {J.}~\bibnamefont {George}},\ }\href
  {https://doi.org/10.26434/chemrxiv.7234721.v4} {\bibfield  {journal}
  {\bibinfo  {journal} {ChemRxiv}\ } (\bibinfo {year} {2019})}\BibitemShut
  {NoStop}%
\bibitem [{\citenamefont {Vergauwe}\ \emph {et~al.}(2019)\citenamefont
  {Vergauwe}, \citenamefont {Thomas}, \citenamefont {Nagarajan}, \citenamefont
  {Shalabney}, \citenamefont {George}, \citenamefont {Chervy}, \citenamefont
  {Seidel}, \citenamefont {Devaux}, \citenamefont {Torbeev},\ and\
  \citenamefont {Ebbesen}}]{Vergauwe2019}%
  \BibitemOpen
  \bibfield  {author} {\bibinfo {author} {\bibfnamefont {R.~M.~A.}\
  \bibnamefont {Vergauwe}}, \bibinfo {author} {\bibfnamefont {A.}~\bibnamefont
  {Thomas}}, \bibinfo {author} {\bibfnamefont {K.}~\bibnamefont {Nagarajan}},
  \bibinfo {author} {\bibfnamefont {A.}~\bibnamefont {Shalabney}}, \bibinfo
  {author} {\bibfnamefont {J.}~\bibnamefont {George}}, \bibinfo {author}
  {\bibfnamefont {T.}~\bibnamefont {Chervy}}, \bibinfo {author} {\bibfnamefont
  {M.}~\bibnamefont {Seidel}}, \bibinfo {author} {\bibfnamefont
  {E.}~\bibnamefont {Devaux}}, \bibinfo {author} {\bibfnamefont
  {V.}~\bibnamefont {Torbeev}},\ and\ \bibinfo {author} {\bibfnamefont {T.~W.}\
  \bibnamefont {Ebbesen}},\ }\href {https://doi.org/10.1002/anie.201908876}
  {\bibfield  {journal} {\bibinfo  {journal} {Angew. Chem., Int. Ed.}\ }\textbf
  {\bibinfo {volume} {58}},\ \bibinfo {pages} {15324} (\bibinfo {year}
  {2019})}\BibitemShut {NoStop}%
\bibitem [{\citenamefont {Lather}\ and\ \citenamefont
  {George}(2021)}]{Lather2021}%
  \BibitemOpen
  \bibfield  {author} {\bibinfo {author} {\bibfnamefont {J.}~\bibnamefont
  {Lather}}\ and\ \bibinfo {author} {\bibfnamefont {J.}~\bibnamefont
  {George}},\ }\href {https://doi.org/10.1021/acs.jpclett.0c03003} {\bibfield
  {journal} {\bibinfo  {journal} {J. Phys. Chem. Lett.}\ }\textbf {\bibinfo
  {volume} {12}},\ \bibinfo {pages} {379} (\bibinfo {year} {2021})}\BibitemShut
  {NoStop}%
\bibitem [{\citenamefont {Hirai}\ \emph
  {et~al.}(2020{\natexlab{c}})\citenamefont {Hirai}, \citenamefont {Ishikawa},
  \citenamefont {Hutchison},\ and\ \citenamefont {Uji-i}}]{Hirai2020chemRxiv}%
  \BibitemOpen
  \bibfield  {author} {\bibinfo {author} {\bibfnamefont {K.}~\bibnamefont
  {Hirai}}, \bibinfo {author} {\bibfnamefont {H.}~\bibnamefont {Ishikawa}},
  \bibinfo {author} {\bibfnamefont {J.}~\bibnamefont {Hutchison}},\ and\
  \bibinfo {author} {\bibfnamefont {H.}~\bibnamefont {Uji-i}},\ }\href
  {https://chemrxiv.org/articles/preprint/Selective_Crystallization_via_Vibrational_Strong_Coupling/13191617/1}
  {\bibfield  {journal} {\bibinfo  {journal} {ChemRxiv}\ } (\bibinfo {year}
  {2020}{\natexlab{c}})}\BibitemShut {NoStop}%
\bibitem [{\citenamefont {Galego}\ \emph {et~al.}(2019)\citenamefont {Galego},
  \citenamefont {Climent}, \citenamefont {Garcia-Vidal},\ and\ \citenamefont
  {Feist}}]{Galego2019}%
  \BibitemOpen
  \bibfield  {author} {\bibinfo {author} {\bibfnamefont {J.}~\bibnamefont
  {Galego}}, \bibinfo {author} {\bibfnamefont {C.}~\bibnamefont {Climent}},
  \bibinfo {author} {\bibfnamefont {F.~J.}\ \bibnamefont {Garcia-Vidal}},\ and\
  \bibinfo {author} {\bibfnamefont {J.}~\bibnamefont {Feist}},\ }\href
  {https://doi.org/10.1103/PhysRevX.9.021057} {\bibfield  {journal} {\bibinfo
  {journal} {Phys. Rev. X}\ }\textbf {\bibinfo {volume} {9}},\ \bibinfo {pages}
  {021057} (\bibinfo {year} {2019})}\BibitemShut {NoStop}%
\bibitem [{\citenamefont {Campos-Gonzalez-Angulo}\ and\ \citenamefont
  {Yuen-Zhou}(2020)}]{Campos-Gonzalez-Angulo2020}%
  \BibitemOpen
  \bibfield  {author} {\bibinfo {author} {\bibfnamefont {J.~A.}\ \bibnamefont
  {Campos-Gonzalez-Angulo}}\ and\ \bibinfo {author} {\bibfnamefont
  {J.}~\bibnamefont {Yuen-Zhou}},\ }\href {https://doi.org/10.1063/5.0007547}
  {\bibfield  {journal} {\bibinfo  {journal} {J. Chem. Phys.}\ }\textbf
  {\bibinfo {volume} {152}},\ \bibinfo {pages} {161101} (\bibinfo {year}
  {2020})}\BibitemShut {NoStop}%
\bibitem [{\citenamefont {Li}\ \emph {et~al.}(2020)\citenamefont {Li},
  \citenamefont {Nitzan},\ and\ \citenamefont {Subotnik}}]{Li2020}%
  \BibitemOpen
  \bibfield  {author} {\bibinfo {author} {\bibfnamefont {T.~E.}\ \bibnamefont
  {Li}}, \bibinfo {author} {\bibfnamefont {A.}~\bibnamefont {Nitzan}},\ and\
  \bibinfo {author} {\bibfnamefont {J.~E.}\ \bibnamefont {Subotnik}},\ }\href
  {https://doi.org/10.1063/5.0006472} {\bibfield  {journal} {\bibinfo
  {journal} {J. Chem. Phys.}\ }\textbf {\bibinfo {volume} {152}},\ \bibinfo
  {pages} {234107} (\bibinfo {year} {2020})}\BibitemShut {NoStop}%
\bibitem [{\citenamefont {Zhdanov}(2020)}]{Zhdanov2020}%
  \BibitemOpen
  \bibfield  {author} {\bibinfo {author} {\bibfnamefont {V.~P.}\ \bibnamefont
  {Zhdanov}},\ }\href
  {https://doi.org/https://doi.org/10.1016/j.chemphys.2020.110767} {\bibfield
  {journal} {\bibinfo  {journal} {Chem. Phys.}\ }\textbf {\bibinfo {volume}
  {535}},\ \bibinfo {pages} {110767} (\bibinfo {year} {2020})}\BibitemShut
  {NoStop}%
\bibitem [{\citenamefont {H\"anggi}\ \emph {et~al.}(1990)\citenamefont
  {H\"anggi}, \citenamefont {Talkner},\ and\ \citenamefont
  {Borkovec}}]{Hanggi1990rev}%
  \BibitemOpen
  \bibfield  {author} {\bibinfo {author} {\bibfnamefont {P.}~\bibnamefont
  {H\"anggi}}, \bibinfo {author} {\bibfnamefont {P.}~\bibnamefont {Talkner}},\
  and\ \bibinfo {author} {\bibfnamefont {M.}~\bibnamefont {Borkovec}},\ }\href
  {https://doi.org/10.1103/RevModPhys.62.251} {\bibfield  {journal} {\bibinfo
  {journal} {Rev. Mod. Phys.}\ }\textbf {\bibinfo {volume} {62}},\ \bibinfo
  {pages} {251} (\bibinfo {year} {1990})}\BibitemShut {NoStop}%
\bibitem [{\citenamefont {Li}\ \emph {et~al.}(2021)\citenamefont {Li},
  \citenamefont {Mandal},\ and\ \citenamefont {Huo}}]{Li2021}%
  \BibitemOpen
  \bibfield  {author} {\bibinfo {author} {\bibfnamefont {X.}~\bibnamefont
  {Li}}, \bibinfo {author} {\bibfnamefont {A.}~\bibnamefont {Mandal}},\ and\
  \bibinfo {author} {\bibfnamefont {P.}~\bibnamefont {Huo}},\ }\href
  {https://doi.org/10.1038/s41467-021-21610-9} {\bibfield  {journal} {\bibinfo
  {journal} {Nat. Commun.}\ }\textbf {\bibinfo {volume} {12}},\ \bibinfo
  {pages} {1315} (\bibinfo {year} {2021})}\BibitemShut {NoStop}%
\bibitem [{\citenamefont {Du}\ \emph {et~al.}(2021)\citenamefont {Du},
  \citenamefont {Campos-Gonzalez-Angulo},\ and\ \citenamefont
  {Yuen-Zhou}}]{Du2021}%
  \BibitemOpen
  \bibfield  {author} {\bibinfo {author} {\bibfnamefont {M.}~\bibnamefont
  {Du}}, \bibinfo {author} {\bibfnamefont {J.~A.}\ \bibnamefont
  {Campos-Gonzalez-Angulo}},\ and\ \bibinfo {author} {\bibfnamefont
  {J.}~\bibnamefont {Yuen-Zhou}},\ }\href {https://doi.org/10.1063/5.0037905}
  {\bibfield  {journal} {\bibinfo  {journal} {J. Chem. Phys.}\ }\textbf
  {\bibinfo {volume} {154}},\ \bibinfo {pages} {084108} (\bibinfo {year}
  {2021})}\BibitemShut {NoStop}%
\bibitem [{\citenamefont {Campos-Gonzalez-Angulo}\ \emph
  {et~al.}(2019)\citenamefont {Campos-Gonzalez-Angulo}, \citenamefont
  {Ribeiro},\ and\ \citenamefont {Yuen-Zhou}}]{Campos-Gonzalez-Angulo2019}%
  \BibitemOpen
  \bibfield  {author} {\bibinfo {author} {\bibfnamefont {J.~A.}\ \bibnamefont
  {Campos-Gonzalez-Angulo}}, \bibinfo {author} {\bibfnamefont {R.~F.}\
  \bibnamefont {Ribeiro}},\ and\ \bibinfo {author} {\bibfnamefont
  {J.}~\bibnamefont {Yuen-Zhou}},\ }\href
  {https://doi.org/10.1038/s41467-019-12636-1} {\bibfield  {journal} {\bibinfo
  {journal} {Nat. Commun.}\ }\textbf {\bibinfo {volume} {10}},\ \bibinfo
  {pages} {4685} (\bibinfo {year} {2019})}\BibitemShut {NoStop}%
\bibitem [{\citenamefont {Phuc}\ \emph {et~al.}(2020)\citenamefont {Phuc},
  \citenamefont {Trung},\ and\ \citenamefont {Ishizaki}}]{Phuc2020}%
  \BibitemOpen
  \bibfield  {author} {\bibinfo {author} {\bibfnamefont {N.~T.}\ \bibnamefont
  {Phuc}}, \bibinfo {author} {\bibfnamefont {P.~Q.}\ \bibnamefont {Trung}},\
  and\ \bibinfo {author} {\bibfnamefont {A.}~\bibnamefont {Ishizaki}},\ }\href
  {https://doi.org/10.1038/s41598-020-62899-8} {\bibfield  {journal} {\bibinfo
  {journal} {Sci. Rep.}\ }\textbf {\bibinfo {volume} {10}},\ \bibinfo {pages}
  {7318} (\bibinfo {year} {2020})}\BibitemShut {NoStop}%
\bibitem [{\citenamefont {Fischer}\ and\ \citenamefont
  {Saalfrank}(2021)}]{Fischer2021}%
  \BibitemOpen
  \bibfield  {author} {\bibinfo {author} {\bibfnamefont {E.~W.}\ \bibnamefont
  {Fischer}}\ and\ \bibinfo {author} {\bibfnamefont {P.}~\bibnamefont
  {Saalfrank}},\ }\href {https://doi.org/10.1063/5.0040853} {\bibfield
  {journal} {\bibinfo  {journal} {J. Chem. Phys.}\ }\textbf {\bibinfo {volume}
  {154}},\ \bibinfo {pages} {104311} (\bibinfo {year} {2021})}\BibitemShut
  {NoStop}%
\bibitem [{\citenamefont {Vurgaftman}\ \emph {et~al.}(2020)\citenamefont
  {Vurgaftman}, \citenamefont {Simpkins}, \citenamefont {Dunkelberger},\ and\
  \citenamefont {Owrutsky}}]{Vurgaftman2020}%
  \BibitemOpen
  \bibfield  {author} {\bibinfo {author} {\bibfnamefont {I.}~\bibnamefont
  {Vurgaftman}}, \bibinfo {author} {\bibfnamefont {B.~S.}\ \bibnamefont
  {Simpkins}}, \bibinfo {author} {\bibfnamefont {A.~D.}\ \bibnamefont
  {Dunkelberger}},\ and\ \bibinfo {author} {\bibfnamefont {J.~C.}\ \bibnamefont
  {Owrutsky}},\ }\href {https://doi.org/10.1021/acs.jpclett.0c00841} {\bibfield
   {journal} {\bibinfo  {journal} {J. Phys. Chem. Lett.}\ }\textbf {\bibinfo
  {volume} {11}},\ \bibinfo {pages} {3557} (\bibinfo {year}
  {2020})}\BibitemShut {NoStop}%
\bibitem [{\citenamefont {Botzung}\ \emph {et~al.}(2020)\citenamefont
  {Botzung}, \citenamefont {Hagenm\"uller}, \citenamefont {Sch\"utz},
  \citenamefont {Dubail}, \citenamefont {Pupillo},\ and\ \citenamefont
  {Schachenmayer}}]{Botzung2020}%
  \BibitemOpen
  \bibfield  {author} {\bibinfo {author} {\bibfnamefont {T.}~\bibnamefont
  {Botzung}}, \bibinfo {author} {\bibfnamefont {D.}~\bibnamefont
  {Hagenm\"uller}}, \bibinfo {author} {\bibfnamefont {S.}~\bibnamefont
  {Sch\"utz}}, \bibinfo {author} {\bibfnamefont {J.}~\bibnamefont {Dubail}},
  \bibinfo {author} {\bibfnamefont {G.}~\bibnamefont {Pupillo}},\ and\ \bibinfo
  {author} {\bibfnamefont {J.}~\bibnamefont {Schachenmayer}},\ }\href
  {https://doi.org/10.1103/PhysRevB.102.144202} {\bibfield  {journal} {\bibinfo
   {journal} {Phys. Rev. B}\ }\textbf {\bibinfo {volume} {102}},\ \bibinfo
  {pages} {144202} (\bibinfo {year} {2020})}\BibitemShut {NoStop}%
\bibitem [{\citenamefont {Scholes}(2020)}]{Scholes2020prs}%
  \BibitemOpen
  \bibfield  {author} {\bibinfo {author} {\bibfnamefont {G.~D.}\ \bibnamefont
  {Scholes}},\ }\href {https://doi.org/10.1098/rspa.2020.0278} {\bibfield
  {journal} {\bibinfo  {journal} {Proc. R. Soc. A}\ }\textbf {\bibinfo {volume}
  {476}},\ \bibinfo {pages} {20200278} (\bibinfo {year} {2020})}\BibitemShut
  {NoStop}%
\bibitem [{\citenamefont {Ch\'avez}\ \emph {et~al.}(2021)\citenamefont
  {Ch\'avez}, \citenamefont {Mattiotti}, \citenamefont {M\'endez-Berm\'udez},
  \citenamefont {Borgonovi},\ and\ \citenamefont {Celardo}}]{Chavez2021}%
  \BibitemOpen
  \bibfield  {author} {\bibinfo {author} {\bibfnamefont {N.~C.}\ \bibnamefont
  {Ch\'avez}}, \bibinfo {author} {\bibfnamefont {F.}~\bibnamefont {Mattiotti}},
  \bibinfo {author} {\bibfnamefont {J.~A.}\ \bibnamefont
  {M\'endez-Berm\'udez}}, \bibinfo {author} {\bibfnamefont {F.}~\bibnamefont
  {Borgonovi}},\ and\ \bibinfo {author} {\bibfnamefont {G.~L.}\ \bibnamefont
  {Celardo}},\ }\href {https://doi.org/10.1103/PhysRevLett.126.153201}
  {\bibfield  {journal} {\bibinfo  {journal} {Phys. Rev. Lett.}\ }\textbf
  {\bibinfo {volume} {126}},\ \bibinfo {pages} {153201} (\bibinfo {year}
  {2021})}\BibitemShut {NoStop}%
\bibitem [{\citenamefont {Sidler}\ \emph {et~al.}(2021)\citenamefont {Sidler},
  \citenamefont {Sch\"afer}, \citenamefont {Ruggenthaler},\ and\ \citenamefont
  {Rubio}}]{Sidler2021}%
  \BibitemOpen
  \bibfield  {author} {\bibinfo {author} {\bibfnamefont {D.}~\bibnamefont
  {Sidler}}, \bibinfo {author} {\bibfnamefont {C.}~\bibnamefont {Sch\"afer}},
  \bibinfo {author} {\bibfnamefont {M.}~\bibnamefont {Ruggenthaler}},\ and\
  \bibinfo {author} {\bibfnamefont {A.}~\bibnamefont {Rubio}},\ }\href
  {https://doi.org/10.1021/acs.jpclett.0c03436} {\bibfield  {journal} {\bibinfo
   {journal} {J. Phys. Chem. Lett.}\ }\textbf {\bibinfo {volume} {12}},\
  \bibinfo {pages} {508} (\bibinfo {year} {2021})}\BibitemShut {NoStop}%
\bibitem [{\citenamefont {Houdr\'e}\ \emph {et~al.}(1996)\citenamefont
  {Houdr\'e}, \citenamefont {Stanley},\ and\ \citenamefont
  {Ilegems}}]{Houdre1996}%
  \BibitemOpen
  \bibfield  {author} {\bibinfo {author} {\bibfnamefont {R.}~\bibnamefont
  {Houdr\'e}}, \bibinfo {author} {\bibfnamefont {R.~P.}\ \bibnamefont
  {Stanley}},\ and\ \bibinfo {author} {\bibfnamefont {M.}~\bibnamefont
  {Ilegems}},\ }\href {https://doi.org/10.1103/PhysRevA.53.2711} {\bibfield
  {journal} {\bibinfo  {journal} {Phys. Rev. A}\ }\textbf {\bibinfo {volume}
  {53}},\ \bibinfo {pages} {2711} (\bibinfo {year} {1996})}\BibitemShut
  {NoStop}%
\bibitem [{\citenamefont {Galego}\ \emph {et~al.}(2017)\citenamefont {Galego},
  \citenamefont {Garcia-Vidal},\ and\ \citenamefont {Feist}}]{Galego2017}%
  \BibitemOpen
  \bibfield  {author} {\bibinfo {author} {\bibfnamefont {J.}~\bibnamefont
  {Galego}}, \bibinfo {author} {\bibfnamefont {F.~J.}\ \bibnamefont
  {Garcia-Vidal}},\ and\ \bibinfo {author} {\bibfnamefont {J.}~\bibnamefont
  {Feist}},\ }\href {https://link.aps.org/doi/10.1103/PhysRevLett.119.136001}
  {\bibfield  {journal} {\bibinfo  {journal} {Phys. Rev. Lett.}\ }\textbf
  {\bibinfo {volume} {119}},\ \bibinfo {pages} {136001} (\bibinfo {year}
  {2017})}\BibitemShut {NoStop}%
\bibitem [{\citenamefont {Kramer}\ and\ \citenamefont
  {MacKinnon}(1993)}]{Kramer1993rev}%
  \BibitemOpen
  \bibfield  {author} {\bibinfo {author} {\bibfnamefont {B.}~\bibnamefont
  {Kramer}}\ and\ \bibinfo {author} {\bibfnamefont {A.}~\bibnamefont
  {MacKinnon}},\ }\href {https://doi.org/10.1088/0034-4885/56/12/001}
  {\bibfield  {journal} {\bibinfo  {journal} {Rep. Prog. Phys.}\ }\textbf
  {\bibinfo {volume} {56}},\ \bibinfo {pages} {1469} (\bibinfo {year}
  {1993})}\BibitemShut {NoStop}%
\bibitem [{\citenamefont {Lather}\ \emph {et~al.}(2019)\citenamefont {Lather},
  \citenamefont {Bhatt}, \citenamefont {Thomas}, \citenamefont {Ebbesen},\ and\
  \citenamefont {George}}]{Lather2019}%
  \BibitemOpen
  \bibfield  {author} {\bibinfo {author} {\bibfnamefont {J.}~\bibnamefont
  {Lather}}, \bibinfo {author} {\bibfnamefont {P.}~\bibnamefont {Bhatt}},
  \bibinfo {author} {\bibfnamefont {A.}~\bibnamefont {Thomas}}, \bibinfo
  {author} {\bibfnamefont {T.~W.}\ \bibnamefont {Ebbesen}},\ and\ \bibinfo
  {author} {\bibfnamefont {J.}~\bibnamefont {George}},\ }\href
  {https://doi.org/10.1002/anie.201905407} {\bibfield  {journal} {\bibinfo
  {journal} {Angew. Chem., Int. Ed.}\ }\textbf {\bibinfo {volume} {58}},\
  \bibinfo {pages} {10635} (\bibinfo {year} {2019})}\BibitemShut {NoStop}%
\bibitem [{\citenamefont {Sch\"utz}\ \emph {et~al.}(2020)\citenamefont
  {Sch\"utz}, \citenamefont {Schachenmayer}, \citenamefont {Hagenm\"uller},
  \citenamefont {Brennen}, \citenamefont {Volz}, \citenamefont {Sandoghdar},
  \citenamefont {Ebbesen}, \citenamefont {Genes},\ and\ \citenamefont
  {Pupillo}}]{Schutz2020}%
  \BibitemOpen
  \bibfield  {author} {\bibinfo {author} {\bibfnamefont {S.}~\bibnamefont
  {Sch\"utz}}, \bibinfo {author} {\bibfnamefont {J.}~\bibnamefont
  {Schachenmayer}}, \bibinfo {author} {\bibfnamefont {D.}~\bibnamefont
  {Hagenm\"uller}}, \bibinfo {author} {\bibfnamefont {G.~K.}\ \bibnamefont
  {Brennen}}, \bibinfo {author} {\bibfnamefont {T.}~\bibnamefont {Volz}},
  \bibinfo {author} {\bibfnamefont {V.}~\bibnamefont {Sandoghdar}}, \bibinfo
  {author} {\bibfnamefont {T.~W.}\ \bibnamefont {Ebbesen}}, \bibinfo {author}
  {\bibfnamefont {C.}~\bibnamefont {Genes}},\ and\ \bibinfo {author}
  {\bibfnamefont {G.}~\bibnamefont {Pupillo}},\ }\href
  {https://doi.org/10.1103/PhysRevLett.124.113602} {\bibfield  {journal}
  {\bibinfo  {journal} {Phys. Rev. Lett.}\ }\textbf {\bibinfo {volume} {124}},\
  \bibinfo {pages} {113602} (\bibinfo {year} {2020})}\BibitemShut {NoStop}%
\bibitem [{\citenamefont {Szidarovszky}\ \emph {et~al.}(2020)\citenamefont
  {Szidarovszky}, \citenamefont {Hal\'asz},\ and\ \citenamefont
  {Vib\'ok}}]{Szidarovszky2020}%
  \BibitemOpen
  \bibfield  {author} {\bibinfo {author} {\bibfnamefont {T.}~\bibnamefont
  {Szidarovszky}}, \bibinfo {author} {\bibfnamefont {G.~J.}\ \bibnamefont
  {Hal\'asz}},\ and\ \bibinfo {author} {\bibfnamefont {A.}~\bibnamefont
  {Vib\'ok}},\ }\href {https://doi.org/10.1088/1367-2630/ab8264} {\bibfield
  {journal} {\bibinfo  {journal} {New J. Phys.}\ }\textbf {\bibinfo {volume}
  {22}},\ \bibinfo {pages} {053001} (\bibinfo {year} {2020})}\BibitemShut
  {NoStop}%
\bibitem [{\citenamefont {Davidsson}\ and\ \citenamefont
  {Kowalewski}(2020)}]{Davidsson2020}%
  \BibitemOpen
  \bibfield  {author} {\bibinfo {author} {\bibfnamefont {E.}~\bibnamefont
  {Davidsson}}\ and\ \bibinfo {author} {\bibfnamefont {M.}~\bibnamefont
  {Kowalewski}},\ }\href {https://doi.org/10.1021/acs.jpca.0c03867} {\bibfield
  {journal} {\bibinfo  {journal} {J. Phys. Chem. A}\ }\textbf {\bibinfo
  {volume} {124}},\ \bibinfo {pages} {4672} (\bibinfo {year}
  {2020})}\BibitemShut {NoStop}%
\bibitem [{SM()}]{SM}%
  \BibitemOpen
  \href@noop {} {}\bibinfo {note} {See Supplemental Material at [URL will be
  inserted by publisher] for Hamiltonian $H_s^{(l)}$, kinetic model and
  discussion of its validity, parameters and numerical methods for calculating
  reaction rates, procedure to determine thermodynamic parameters of
  activation, derivations of analytical reaction rates, calculations using
  different inhomogeneous broadening, and supplemental figures, as well as
  Refs. [62-76].}\BibitemShut {Stop}%
\bibitem [{\citenamefont {Marcus}(1964)}]{Marcus1964}%
  \BibitemOpen
  \bibfield  {author} {\bibinfo {author} {\bibfnamefont {R.~A.}\ \bibnamefont
  {Marcus}},\ }\href {https://doi.org/10.1146/annurev.pc.15.100164.001103}
  {\bibfield  {journal} {\bibinfo  {journal} {Annu. Rev. Phys. Chem.}\ }\textbf
  {\bibinfo {volume} {15}},\ \bibinfo {pages} {155} (\bibinfo {year}
  {1964})}\BibitemShut {NoStop}%
\bibitem [{\citenamefont {Levich}(1966)}]{Levich1966}%
  \BibitemOpen
  \bibfield  {author} {\bibinfo {author} {\bibfnamefont {V.~G.}\ \bibnamefont
  {Levich}},\ }\href@noop {} {\bibfield  {journal} {\bibinfo  {journal} {Adv.
  Electrochem. Electrochem. Eng}\ }\textbf {\bibinfo {volume} {4}},\ \bibinfo
  {pages} {249} (\bibinfo {year} {1966})}\BibitemShut {NoStop}%
\bibitem [{\citenamefont {Jortner}(1976)}]{Jortner1976}%
  \BibitemOpen
  \bibfield  {author} {\bibinfo {author} {\bibfnamefont {J.}~\bibnamefont
  {Jortner}},\ }\href {https://doi.org/10.1063/1.432142} {\bibfield  {journal}
  {\bibinfo  {journal} {J. Chem. Phys.}\ }\textbf {\bibinfo {volume} {64}},\
  \bibinfo {pages} {4860} (\bibinfo {year} {1976})}\BibitemShut {NoStop}%
\bibitem [{\citenamefont {May}\ and\ \citenamefont {K\"{u}hn}(2011)}]{MayBook}%
  \BibitemOpen
  \bibfield  {author} {\bibinfo {author} {\bibfnamefont {V.}~\bibnamefont
  {May}}\ and\ \bibinfo {author} {\bibfnamefont {O.}~\bibnamefont {K\"{u}hn}},\
  }\href@noop {} {\emph {\bibinfo {title} {Charge and Energy Transfer Dynamics
  in Molecular Systems}}}\ (\bibinfo  {publisher} {Wiley},\ \bibinfo {year}
  {2011})\BibitemShut {NoStop}%
\bibitem [{\citenamefont {del Pino}\ \emph {et~al.}(2015)\citenamefont {del
  Pino}, \citenamefont {Feist},\ and\ \citenamefont
  {Garcia-Vidal}}]{delPino2015}%
  \BibitemOpen
  \bibfield  {author} {\bibinfo {author} {\bibfnamefont {J.}~\bibnamefont {del
  Pino}}, \bibinfo {author} {\bibfnamefont {J.}~\bibnamefont {Feist}},\ and\
  \bibinfo {author} {\bibfnamefont {F.~J.}\ \bibnamefont {Garcia-Vidal}},\
  }\href {http://stacks.iop.org/1367-2630/17/i=5/a=053040} {\bibfield
  {journal} {\bibinfo  {journal} {New J. Phys.}\ }\textbf {\bibinfo {volume}
  {17}},\ \bibinfo {pages} {053040} (\bibinfo {year} {2015})}\BibitemShut
  {NoStop}%
\bibitem [{\citenamefont {Xiang}\ \emph {et~al.}(2018)\citenamefont {Xiang},
  \citenamefont {Ribeiro}, \citenamefont {Dunkelberger}, \citenamefont {Wang},
  \citenamefont {Li}, \citenamefont {Simpkins}, \citenamefont {Owrutsky},
  \citenamefont {Yuen-Zhou},\ and\ \citenamefont {Xiong}}]{Xiang2018}%
  \BibitemOpen
  \bibfield  {author} {\bibinfo {author} {\bibfnamefont {B.}~\bibnamefont
  {Xiang}}, \bibinfo {author} {\bibfnamefont {R.~F.}\ \bibnamefont {Ribeiro}},
  \bibinfo {author} {\bibfnamefont {A.~D.}\ \bibnamefont {Dunkelberger}},
  \bibinfo {author} {\bibfnamefont {J.}~\bibnamefont {Wang}}, \bibinfo {author}
  {\bibfnamefont {Y.}~\bibnamefont {Li}}, \bibinfo {author} {\bibfnamefont
  {B.~S.}\ \bibnamefont {Simpkins}}, \bibinfo {author} {\bibfnamefont {J.~C.}\
  \bibnamefont {Owrutsky}}, \bibinfo {author} {\bibfnamefont {J.}~\bibnamefont
  {Yuen-Zhou}},\ and\ \bibinfo {author} {\bibfnamefont {W.}~\bibnamefont
  {Xiong}},\ }\href {https://doi.org/10.1073/pnas.1722063115} {\bibfield
  {journal} {\bibinfo  {journal} {Proc. Natl. Acad. Sci. U. S. A.}\ }\textbf
  {\bibinfo {volume} {115}},\ \bibinfo {pages} {4845} (\bibinfo {year}
  {2018})}\BibitemShut {NoStop}%
\bibitem [{\citenamefont {Xiang}\ \emph
  {et~al.}(2019{\natexlab{a}})\citenamefont {Xiang}, \citenamefont {Ribeiro},
  \citenamefont {Chen}, \citenamefont {Wang}, \citenamefont {Du}, \citenamefont
  {Yuen-Zhou},\ and\ \citenamefont {Xiong}}]{Xiang2019jpca}%
  \BibitemOpen
  \bibfield  {author} {\bibinfo {author} {\bibfnamefont {B.}~\bibnamefont
  {Xiang}}, \bibinfo {author} {\bibfnamefont {R.~F.}\ \bibnamefont {Ribeiro}},
  \bibinfo {author} {\bibfnamefont {L.}~\bibnamefont {Chen}}, \bibinfo {author}
  {\bibfnamefont {J.}~\bibnamefont {Wang}}, \bibinfo {author} {\bibfnamefont
  {M.}~\bibnamefont {Du}}, \bibinfo {author} {\bibfnamefont {J.}~\bibnamefont
  {Yuen-Zhou}},\ and\ \bibinfo {author} {\bibfnamefont {W.}~\bibnamefont
  {Xiong}},\ }\href {https://doi.org/10.1021/acs.jpca.9b04601} {\bibfield
  {journal} {\bibinfo  {journal} {J. Phys. Chem. A}\ }\textbf {\bibinfo
  {volume} {123}},\ \bibinfo {pages} {5918} (\bibinfo {year}
  {2019}{\natexlab{a}})}\BibitemShut {NoStop}%
\bibitem [{\citenamefont {Ribeiro}\ \emph
  {et~al.}(2018{\natexlab{b}})\citenamefont {Ribeiro}, \citenamefont
  {Dunkelberger}, \citenamefont {Xiang}, \citenamefont {Xiong}, \citenamefont
  {Simpkins}, \citenamefont {Owrutsky},\ and\ \citenamefont
  {Yuen-Zhou}}]{Ribeiro2018vp}%
  \BibitemOpen
  \bibfield  {author} {\bibinfo {author} {\bibfnamefont {R.~F.}\ \bibnamefont
  {Ribeiro}}, \bibinfo {author} {\bibfnamefont {A.~D.}\ \bibnamefont
  {Dunkelberger}}, \bibinfo {author} {\bibfnamefont {B.}~\bibnamefont {Xiang}},
  \bibinfo {author} {\bibfnamefont {W.}~\bibnamefont {Xiong}}, \bibinfo
  {author} {\bibfnamefont {B.~S.}\ \bibnamefont {Simpkins}}, \bibinfo {author}
  {\bibfnamefont {J.~C.}\ \bibnamefont {Owrutsky}},\ and\ \bibinfo {author}
  {\bibfnamefont {J.}~\bibnamefont {Yuen-Zhou}},\ }\href
  {https://doi.org/10.1021/acs.jpclett.8b01176} {\bibfield  {journal} {\bibinfo
   {journal} {J. Phys. Chem. Lett.}\ }\textbf {\bibinfo {volume} {9}},\
  \bibinfo {pages} {3766} (\bibinfo {year} {2018}{\natexlab{b}})}\BibitemShut
  {NoStop}%
\bibitem [{\citenamefont {Canales}\ \emph {et~al.}(2021)\citenamefont
  {Canales}, \citenamefont {Baranov}, \citenamefont {Antosiewicz},\ and\
  \citenamefont {Shegai}}]{Canales2021}%
  \BibitemOpen
  \bibfield  {author} {\bibinfo {author} {\bibfnamefont {A.}~\bibnamefont
  {Canales}}, \bibinfo {author} {\bibfnamefont {D.~G.}\ \bibnamefont
  {Baranov}}, \bibinfo {author} {\bibfnamefont {T.~J.}\ \bibnamefont
  {Antosiewicz}},\ and\ \bibinfo {author} {\bibfnamefont {T.}~\bibnamefont
  {Shegai}},\ }\href {https://doi.org/10.1063/5.0033352} {\bibfield  {journal}
  {\bibinfo  {journal} {J. Chem. Phys.}\ }\textbf {\bibinfo {volume} {154}},\
  \bibinfo {pages} {024701} (\bibinfo {year} {2021})}\BibitemShut {NoStop}%
\bibitem [{\citenamefont {Agarwal}(2013)}]{AgarwalBook}%
  \BibitemOpen
  \bibfield  {author} {\bibinfo {author} {\bibfnamefont {G.}~\bibnamefont
  {Agarwal}},\ }\href@noop {} {\emph {\bibinfo {title} {Quantum Optics}}}\
  (\bibinfo  {publisher} {Cambridge University Press},\ \bibinfo {year}
  {2013})\BibitemShut {NoStop}%
\bibitem [{\citenamefont {Agranovich}\ \emph {et~al.}(2003)\citenamefont
  {Agranovich}, \citenamefont {Litinskaia},\ and\ \citenamefont
  {Lidzey}}]{Agranovich2003}%
  \BibitemOpen
  \bibfield  {author} {\bibinfo {author} {\bibfnamefont {V.~M.}\ \bibnamefont
  {Agranovich}}, \bibinfo {author} {\bibfnamefont {M.}~\bibnamefont
  {Litinskaia}},\ and\ \bibinfo {author} {\bibfnamefont {D.~G.}\ \bibnamefont
  {Lidzey}},\ }\href {https://link.aps.org/doi/10.1103/PhysRevB.67.085311}
  {\bibfield  {journal} {\bibinfo  {journal} {Phys. Rev. B}\ }\textbf {\bibinfo
  {volume} {67}},\ \bibinfo {pages} {085311} (\bibinfo {year}
  {2003})}\BibitemShut {NoStop}%
\bibitem [{\citenamefont {Litinskaya}\ \emph {et~al.}(2004)\citenamefont
  {Litinskaya}, \citenamefont {Reineker},\ and\ \citenamefont
  {Agranovich}}]{Litinskaya2004}%
  \BibitemOpen
  \bibfield  {author} {\bibinfo {author} {\bibfnamefont {M.}~\bibnamefont
  {Litinskaya}}, \bibinfo {author} {\bibfnamefont {P.}~\bibnamefont
  {Reineker}},\ and\ \bibinfo {author} {\bibfnamefont {V.~M.}\ \bibnamefont
  {Agranovich}},\ }\href
  {https://doi.org/https://doi.org/10.1016/j.jlumin.2004.08.033} {\bibfield
  {journal} {\bibinfo  {journal} {J. Lumin.}\ }\textbf {\bibinfo {volume}
  {110}},\ \bibinfo {pages} {364} (\bibinfo {year} {2004})}\BibitemShut
  {NoStop}%
\bibitem [{\citenamefont {Dunkelberger}\ \emph {et~al.}(2016)\citenamefont
  {Dunkelberger}, \citenamefont {Spann}, \citenamefont {Fears}, \citenamefont
  {Simpkins},\ and\ \citenamefont {Owrutsky}}]{Dunkelberger2016}%
  \BibitemOpen
  \bibfield  {author} {\bibinfo {author} {\bibfnamefont {A.~D.}\ \bibnamefont
  {Dunkelberger}}, \bibinfo {author} {\bibfnamefont {B.~T.}\ \bibnamefont
  {Spann}}, \bibinfo {author} {\bibfnamefont {K.~P.}\ \bibnamefont {Fears}},
  \bibinfo {author} {\bibfnamefont {B.~S.}\ \bibnamefont {Simpkins}},\ and\
  \bibinfo {author} {\bibfnamefont {J.~C.}\ \bibnamefont {Owrutsky}},\ }\href
  {https://doi.org/10.1038/ncomms13504
  https://www.nature.com/articles/ncomms13504#supplementary-information}
  {\bibfield  {journal} {\bibinfo  {journal} {Nat. Commun.}\ }\textbf {\bibinfo
  {volume} {7}},\ \bibinfo {pages} {13504} (\bibinfo {year}
  {2016})}\BibitemShut {NoStop}%
\bibitem [{\citenamefont {Grafton}\ \emph {et~al.}(2021)\citenamefont
  {Grafton}, \citenamefont {Dunkelberger}, \citenamefont {Simpkins},
  \citenamefont {Triana}, \citenamefont {Hern\'andez}, \citenamefont
  {Herrera},\ and\ \citenamefont {Owrutsky}}]{Grafton2021}%
  \BibitemOpen
  \bibfield  {author} {\bibinfo {author} {\bibfnamefont {A.~B.}\ \bibnamefont
  {Grafton}}, \bibinfo {author} {\bibfnamefont {A.~D.}\ \bibnamefont
  {Dunkelberger}}, \bibinfo {author} {\bibfnamefont {B.~S.}\ \bibnamefont
  {Simpkins}}, \bibinfo {author} {\bibfnamefont {J.~F.}\ \bibnamefont
  {Triana}}, \bibinfo {author} {\bibfnamefont {F.~J.}\ \bibnamefont
  {Hern\'andez}}, \bibinfo {author} {\bibfnamefont {F.}~\bibnamefont
  {Herrera}},\ and\ \bibinfo {author} {\bibfnamefont {J.~C.}\ \bibnamefont
  {Owrutsky}},\ }\href {https://doi.org/10.1038/s41467-020-20535-z} {\bibfield
  {journal} {\bibinfo  {journal} {Nat. Commun.}\ }\textbf {\bibinfo {volume}
  {12}},\ \bibinfo {pages} {214} (\bibinfo {year} {2021})}\BibitemShut
  {NoStop}%
\bibitem [{\citenamefont {Duan}\ \emph {et~al.}(2017)\citenamefont {Duan},
  \citenamefont {Wang}, \citenamefont {Tang},\ and\ \citenamefont
  {Wu}}]{Duan2017}%
  \BibitemOpen
  \bibfield  {author} {\bibinfo {author} {\bibfnamefont {C.}~\bibnamefont
  {Duan}}, \bibinfo {author} {\bibfnamefont {Q.}~\bibnamefont {Wang}}, \bibinfo
  {author} {\bibfnamefont {Z.}~\bibnamefont {Tang}},\ and\ \bibinfo {author}
  {\bibfnamefont {J.}~\bibnamefont {Wu}},\ }\href
  {https://doi.org/10.1063/1.4997669} {\bibfield  {journal} {\bibinfo
  {journal} {J. Chem. Phys.}\ }\textbf {\bibinfo {volume} {147}},\ \bibinfo
  {pages} {164112} (\bibinfo {year} {2017})}\BibitemShut {NoStop}%
\bibitem [{\citenamefont {K\"uhn}\ and\ \citenamefont
  {Naundorf}(2003)}]{Kuhn2003}%
  \BibitemOpen
  \bibfield  {author} {\bibinfo {author} {\bibfnamefont {O.}~\bibnamefont
  {K\"uhn}}\ and\ \bibinfo {author} {\bibfnamefont {H.}~\bibnamefont
  {Naundorf}},\ }\href {https://doi.org/10.1039/B209587D} {\bibfield  {journal}
  {\bibinfo  {journal} {Phys. Chem. Chem. Phys.}\ }\textbf {\bibinfo {volume}
  {5}},\ \bibinfo {pages} {79} (\bibinfo {year} {2003})}\BibitemShut {NoStop}%
\bibitem [{\citenamefont {V\"ohringer}\ \emph {et~al.}(1995)\citenamefont
  {V\"ohringer}, \citenamefont {Arnett}, \citenamefont {Westervelt},
  \citenamefont {Feldstein},\ and\ \citenamefont {Scherer}}]{Vohringer1995}%
  \BibitemOpen
  \bibfield  {author} {\bibinfo {author} {\bibfnamefont {P.}~\bibnamefont
  {V\"ohringer}}, \bibinfo {author} {\bibfnamefont {D.~C.}\ \bibnamefont
  {Arnett}}, \bibinfo {author} {\bibfnamefont {R.~A.}\ \bibnamefont
  {Westervelt}}, \bibinfo {author} {\bibfnamefont {M.~J.}\ \bibnamefont
  {Feldstein}},\ and\ \bibinfo {author} {\bibfnamefont {N.~F.}\ \bibnamefont
  {Scherer}},\ }\href {https://doi.org/10.1063/1.468531} {\bibfield  {journal}
  {\bibinfo  {journal} {J. Chem. Phys.}\ }\textbf {\bibinfo {volume} {102}},\
  \bibinfo {pages} {4027} (\bibinfo {year} {1995})}\BibitemShut {NoStop}%
\bibitem [{\citenamefont {Chang}\ and\ \citenamefont
  {Castner}(1993)}]{Chang1993}%
  \BibitemOpen
  \bibfield  {author} {\bibinfo {author} {\bibfnamefont {Y.~J.}\ \bibnamefont
  {Chang}}\ and\ \bibinfo {author} {\bibfnamefont {E.~W.}\ \bibnamefont
  {Castner}},\ }\href {https://doi.org/10.1063/1.465710} {\bibfield  {journal}
  {\bibinfo  {journal} {J. Chem. Phys.}\ }\textbf {\bibinfo {volume} {99}},\
  \bibinfo {pages} {7289} (\bibinfo {year} {1993})}\BibitemShut {NoStop}%
\bibitem [{\citenamefont {Xiang}\ \emph
  {et~al.}(2019{\natexlab{b}})\citenamefont {Xiang}, \citenamefont {Ribeiro},
  \citenamefont {Li}, \citenamefont {Dunkelberger}, \citenamefont {Simpkins},
  \citenamefont {Yuen-Zhou},\ and\ \citenamefont {Xiong}}]{Xiang2019sa}%
  \BibitemOpen
  \bibfield  {author} {\bibinfo {author} {\bibfnamefont {B.}~\bibnamefont
  {Xiang}}, \bibinfo {author} {\bibfnamefont {R.~F.}\ \bibnamefont {Ribeiro}},
  \bibinfo {author} {\bibfnamefont {Y.}~\bibnamefont {Li}}, \bibinfo {author}
  {\bibfnamefont {A.~D.}\ \bibnamefont {Dunkelberger}}, \bibinfo {author}
  {\bibfnamefont {B.~B.}\ \bibnamefont {Simpkins}}, \bibinfo {author}
  {\bibfnamefont {J.}~\bibnamefont {Yuen-Zhou}},\ and\ \bibinfo {author}
  {\bibfnamefont {W.}~\bibnamefont {Xiong}},\ }\href
  {https://doi.org/10.1126/sciadv.aax5196} {\bibfield  {journal} {\bibinfo
  {journal} {Sci. Adv.}\ }\textbf {\bibinfo {volume} {5}},\ \bibinfo {pages}
  {eaax5196} (\bibinfo {year} {2019}{\natexlab{b}})}\BibitemShut {NoStop}%
\bibitem [{\citenamefont {Breuer}\ and\ \citenamefont
  {Petruccione}(2002)}]{BreuerBook}%
  \BibitemOpen
  \bibfield  {author} {\bibinfo {author} {\bibfnamefont {H.}~\bibnamefont
  {Breuer}}\ and\ \bibinfo {author} {\bibfnamefont {F.}~\bibnamefont
  {Petruccione}},\ }\href@noop {} {\emph {\bibinfo {title} {The Theory of Open
  Quantum Systems}}}\ (\bibinfo  {publisher} {OUP Oxford},\ \bibinfo {year}
  {2002})\BibitemShut {NoStop}%
\bibitem [{\citenamefont {Frankcombe}\ and\ \citenamefont
  {Smith}(2001)}]{Frankcombe2001}%
  \BibitemOpen
  \bibfield  {author} {\bibinfo {author} {\bibfnamefont {T.~J.}\ \bibnamefont
  {Frankcombe}}\ and\ \bibinfo {author} {\bibfnamefont {S.~C.}\ \bibnamefont
  {Smith}},\ }\href
  {https://doi.org/https://doi.org/10.1016/S0010-4655(01)00298-3} {\bibfield
  {journal} {\bibinfo  {journal} {Comput. Phys. Commun.}\ }\textbf {\bibinfo
  {volume} {141}},\ \bibinfo {pages} {39} (\bibinfo {year} {2001})}\BibitemShut
  {NoStop}%
\bibitem [{\citenamefont {Atkins}\ and\ \citenamefont
  {de~Paula}(2010)}]{AtkinsBook9th}%
  \BibitemOpen
  \bibfield  {author} {\bibinfo {author} {\bibfnamefont {P.}~\bibnamefont
  {Atkins}}\ and\ \bibinfo {author} {\bibfnamefont {J.}~\bibnamefont
  {de~Paula}},\ }\href@noop {} {\emph {\bibinfo {title} {Physical
  Chemistry}}},\ \bibinfo {edition} {9th}\ ed.\ (\bibinfo  {publisher} {OUP
  Oxford},\ \bibinfo {year} {2010})\BibitemShut {NoStop}%
\bibitem [{\citenamefont {Eyring}(1935)}]{Eyring1935jcp}%
  \BibitemOpen
  \bibfield  {author} {\bibinfo {author} {\bibfnamefont {H.}~\bibnamefont
  {Eyring}},\ }\href {https://doi.org/10.1063/1.1749604} {\bibfield  {journal}
  {\bibinfo  {journal} {J. Chem. Phys.}\ }\textbf {\bibinfo {volume} {3}},\
  \bibinfo {pages} {107} (\bibinfo {year} {1935})}\BibitemShut {NoStop}%
\bibitem [{\citenamefont {Evans}\ and\ \citenamefont
  {Polanyi}(1935)}]{Evans1935}%
  \BibitemOpen
  \bibfield  {author} {\bibinfo {author} {\bibfnamefont {M.~G.}\ \bibnamefont
  {Evans}}\ and\ \bibinfo {author} {\bibfnamefont {M.}~\bibnamefont
  {Polanyi}},\ }\href {https://doi.org/10.1039/TF9353100875} {\bibfield
  {journal} {\bibinfo  {journal} {Trans. Faraday Soc.}\ }\textbf {\bibinfo
  {volume} {31}},\ \bibinfo {pages} {875} (\bibinfo {year} {1935})}\BibitemShut
  {NoStop}%
\end{thebibliography}

\end{document}


\onecolumngrid
\makeatletter 
\renewcommand\@biblabel[1]{[S#1]}  
\renewcommand{\citenumfont}[1]{S#1} 
\makeatother

\setcounter{figure}{0} \renewcommand{\thefigure}{S\arabic{figure}}
\setcounter{equation}{0} \renewcommand{\theequation}{S\arabic{equation}}

\title{Supplemental Material: Catalysis by Dark States in Vibropolaritonic Chemistry}
\author{Matthew Du, Joel Yuen-Zhou}
\email{joelyuen@ucsd.edu}

\address{Department of Chemistry and Biochemistry, University of California
San Diego, La Jolla, California 92093, United States}
\date{\today}
\maketitle

\section{Hamiltonian $H_{s}^{(l)}$}

The Hamiltonian $H_{s}^{(l)}$, which is the last term of Hamiltonian
$H_{\text{rxn}}$ (defined in main text), describes the low-frequency
vibrational modes of the solvent that help mediate electron transfer.
We define the former Hamiltonian as
\[
H_{s}^{(l)}=\hbar\sum_{j}\omega_{j}^{(l)}b_{j}^{\dagger}b_{j}+\hbar\sum_{j}\omega_{j}^{(l)}\sum_{X=R,P}|X\rangle\langle X|\left[\lambda_{Xj}^{(l)}(b_{j}+b_{j}^{\dagger})+(\lambda_{Xj}^{(l)})^{2}\right],
\]
where the $j$th low-frequency mode has frequency $\omega_{j}^{(l)}$
and couples to electronic state $|X\rangle$ with dimensionless strength
$\lambda_{Xj}^{(l)}$. The contribution of the low-frequency modes to the
reaction rate is characterized by reorganization energy $\lambda_{s}=\sum_{j}\left(\lambda_{Pj}^{(l)}-\lambda_{Rj}^{(l)}\right)^{2}\hbar\omega_{j}^{(l)}$.

\section{kinetic model \label{sec:kinetic-model}}

Here, we describe in detail the kinetic model used to simulate the
reaction. As discussed in the main text, this model only considers
states $|X,\chi\rangle$ whose vibrational-cavity component has zero
($\chi=0$) or one ($\chi=1_{q}$, where $q=1,\dots,N+1$) excitation.
The population $p_{(X,\chi)}$ of state $|X,\chi\rangle$ evolves
according to the master equation
\begin{align}
\frac{dp_{(X,\chi)}}{dt} & =-\sum_{(X',\chi')\neq(X,\chi)}k_{(X,\chi)\rightarrow(X',\chi')}p_{(X,\chi)}+\sum_{(X',\chi')\neq(X,\chi)}k_{(X',\chi')\rightarrow(X,\chi)}p_{(X',\chi')}.\label{eq:master-eq}
\end{align}
State-to-state transitions ($|X,\chi\rangle\rightarrow|X',\chi'\rangle$)
are either reactive ($X'\neq X$) or nonreactive ($X'=X$). 

Rates ($k_{(X,\chi)\rightarrow(X',\chi')}$ for $X'\neq X$) of reactive
transitions are given by Eq. (\ref{eq:k_ET}) and depend on FC factor
$F_{\chi,\chi'}$, which is defined in the main text. We now evaluate
$F_{\chi,\chi'}$ for various combinations of $\chi$ and $\chi'$.
Recall the standard identity \citep{AgarwalBook}
\begin{equation}
\langle m'|D(\lambda)|m\rangle=\sqrt{\frac{m!}{(m')!}}e^{-|\lambda|^{2}/2}\lambda^{m'-m}L_{m}^{(m'-m)}(|\lambda|^{2}),\quad m'\geq m,\label{eq:id_displ}
\end{equation}
where $D(\lambda)=\exp(\lambda a^{\dagger}-\lambda^{*}a)$ is the
displacement operator corresponding to bosonic annihilation operator
$a$, $|m\rangle$ is a number state of the mode represented by $a$,
and $L_{n}^{k}(x)$ is an associated Laguerre polynomial. Using Eq.
(\ref{eq:id_displ}), we obtain
\begin{equation}
F_{\chi,\chi'}=\begin{cases}
e^{-S}, & (\chi',\chi)=(0,0),\\
e^{-S}S_{q}, & (\chi',\chi)=(1_{q},0),(0,1_{q}).\\
e^{-S}(1-S_{q})^{2}, & (\chi',\chi)=(1_{q},1_{q}).\\
e^{-S}S_{q'}S_{q}, & (\chi',\chi)=(1_{q'},1_{q}),\quad q'\neq q,
\end{cases}
\end{equation}
where $S=\sum_{q=1}^{N+1}S_{q}$ and $S_{q}=|\lambda_{Pq}-\lambda_{Rq}|^{2}$. 

The nonreactive transitions in our model are of two types: (1) decay/gain
of an excitation in vibrational-cavity mode $q$ and (2) energy exchange
among dark and polariton states. An excitation in mode $q$ decays
at rate 
\begin{equation}
k_{(X,1_{q})\rightarrow(X,0)}=|c_{q0}|^{2}\kappa+\left(\sum_{i=1}^{N}|c_{qi}|^{2}\right)\gamma,\label{eq:k_decay}
\end{equation}
where $\kappa$ is the decay (leakage) rate of the bare cavity and
$\gamma$ is the decay rate of all bare vibrations. Detailed balance
governs the rate of the reverse process: $k_{(X,0)\rightarrow(X,1_{q})}=k_{(X,1_{q})\rightarrow(X,0)}\exp(-\beta\hbar\omega_{q})$.
Relaxation among dark and polariton states is driven by
vibrational dephasing interactions, i.e., anharmonic
coupling between bare molecular vibrations and their local chemical
environment \citep{delPino2015,Xiang2018,Xiang2019jpca}. Following
theories \citep{delPino2015,Agranovich2003,Litinskaya2004} of relaxation
dynamics for molecules under VSC, the transition from a polariton
or dark state to another has rate
\begin{equation}
k_{(X,1_{q})\rightarrow(X,1_{q'})}=
\left(\sum_{i=1}^{N}|c_{q'i}|^{2}|c_{qi}|^{2}\right)\mathcal{R}(\omega_{q'}-\omega_{q}),
\label{eq:k_redis}
\end{equation}
where $q'\neq q$,
\begin{equation}
\mathcal{R}(\omega)= 
2\pi\left[\Theta(-\omega)\left(n(-\omega)+1\right)J(-\omega)+\Theta(\omega)n(\omega)J(\omega)\right],
\label{eq:r}
\end{equation}
and $\Theta(\omega)$ is the Heaviside step function.
The environmental modes are characterized by spectral density $J(\omega)$
and the Bose-Einstein distribution function, $n(\omega)=\left(\exp(\beta\hbar\omega)-1\right)^{-1}$. 

\section{Default parameters for calculations of reaction rate \label{sec:default-parameters}}

Unless otherwise stated, calculations of reaction rates are carried
out using the parameters described in this section. 

For the bare molecular vibrations, we choose a mean frequency of $\overline{\omega}_{v}=2000$
cm$^{-1}$, which is representative of experimental studies on VSC
\citep{Dunkelberger2016,Xiang2018,Lather2019,Xiang2019jpca,Hirai2020,Grafton2021}.
Since these studies do not report values of inhomogeneous broadening,
we simply take the vibrational frequencies to have a standard deviation
of $\sigma_{v}=10$ cm$^{-1}$. This choice of $\sigma_{v}$ yields
a spectral linewidth of $\approx24$ cm$^{-1}$, which is consistent
with vibrational lineshapes measured in some of the cited works \citep{Lather2019,Xiang2019jpca,Hirai2020,Grafton2021}.
For the cavity frequency and cavity-vibration interaction, we use 
$\omega_c = \overline{\omega}_v$ and $g\sqrt{N}=8\sigma_{v}$ (for all $N$), respectively.
Regarding the electronic degree of freedom and its coupling to other
degrees of freedom, we select parameters employed in the reaction
simulations of \citep{Du2021}: $E_{R}=0$, $E_{P}=-0.6\overline{\omega}_{v}$,
$\lambda_{R}=0$, $\lambda_{P}=1.5$, $J_{RP}=0.01\overline{\omega}_{v}$,
$\lambda_{s}=0.08\overline{\omega}_{v}$. 

Next, we describe the parameters governing the relaxation of vibrational
and cavity modes. The temperature $T$ is set to 298 K. We choose
$\kappa=1$ ps$^{-1}$ as the bare cavity decay rate and $\gamma=0.01$
ps$^{-1}$ for the decay rate of all bare vibrational modes \citep{Xiang2019jpca}.
To model the relaxation among dark and polariton states {[}Eq. (\ref{eq:k_redis}){]},
we use the super-Ohmic spectral density
\begin{equation}
J(\omega)=\eta\omega_0^{-1}\omega^2\exp[-(\omega/\omega_{\text{cut}})^2],
\label{eq:j}
\end{equation}
where $\eta=1\times10^{-3}$ is the interaction strength between each
bare vibrational mode and its local chemical environment, $\omega_{\text{cut}}=50$
cm$^{-1}$ is the cutoff frequency of the environmental modes, 
and $\omega_0=\omega_{\text{cut}}$. The
spectral density resembles those in models of 
condensed-phase systems in general \citep{Duan2017},
while the cutoff frequency is similar to those in
liquid-phase molecular systems
\citep{Kuhn2003,Vohringer1995,Chang1993}.

\begin{figure}[H]
\centering\includegraphics{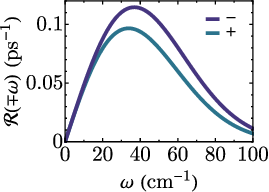}
	
\caption{\label{fig:r}$\mathcal{R}(\mp\omega)$ [Eq. (\ref{eq:r})] for 
$J(\omega)$ of Eq. (\ref{eq:j}) and $T=298$ K.
}
	
\end{figure}

Our choice of spectral density is motivated by a number of other factors.  
First, $J(\omega)$ of  Eq. (\ref{eq:j}) allows for relaxation from polaritons to dark states that 
occurs on a timescale of 20-25 ps for $g\sqrt{N}\approx 20$ cm$^{-1}$.
This value is estimated by calculating 
the corresponding rate constant [Eq. (\ref{eq:k_redis})] as
$\mathcal{R}(\mp 20\text{ cm}^{-1})/2$ (Fig. \ref{fig:r}), 
which applies when the initial state is the upper/lower polariton \citep{delPino2015}.
Measurements of vibrational polaritons using ultrafast 2D IR spectroscopy have revealed that,
by 25 ps (but not less than 5 ps) after polaritons are excited, 
the dark states significantly contribute to the transient absorption signal \citep{Xiang2018,Xiang2019sa}.
  
Second, we choose $J(\omega)$ such that the rate of relaxation between 
dark states decreases as the frequency between the states approaches zero 
[see Fig. \ref{fig:r}, namely $\mathcal{R}(\mp\omega)$ as $\omega \rightarrow 0$].
We do this so that the associated decay linewidths 
of dark states do not 
exceed their energy spacing, maintaining the validity of 
the kinetic model in this work (\S\ref{sec:valid-largeN}). 
For the same reason, we set $N\leq32$ throughout this work (\S\ref{sec:valid-largeN}).

\section{Neglect of higher excitation manifolds in the kinetic model} 

Our kinetic model (\S\ref{sec:kinetic-model}) only includes states with 
0 or 1 total excitations in the vibrational-cavity modes (\emph{i.e.}, dark and polariton modes).  
Below, we show that the truncation of higher excitation manifolds is valid for
the default calculation parameters
(\S\ref{sec:default-parameters}).
We also argue that this approximation holds for other parameter sets used in this work.

\begin{figure}[H]
\centering\includegraphics{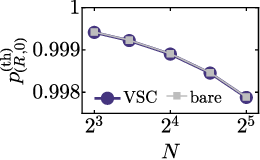}
	
\caption{\label{fig:pop0}Population ($p_{(R,0)}^{(\text{th})}$) of $|R,0\rangle$ for 
a thermal distribution ($T$ = 298 K) of 
reactant eigenstates $|R,\chi\rangle$,
where the states span all excitation manifolds of the vibrational-cavity subspace.
The populations, which are averages over 5000 disorder realizations, 
are shown for various $N$ and regimes of light-matter coupling (VSC, bare).
Other calculation parameters are given in \S\ref{sec:default-parameters}.
}

\end{figure}

Since $E_R$ is closer to $E_P + \hbar\overline{\omega}_v$ than it is to $E_P$,
the major reactive transitions are between reactant states ($|R,\chi\rangle$) with 
$N_\text{ex}$ excitations in the vibrational-cavity modes 
(specifically modes that overlap most with the reactive mode) 
and product states ($|P,\chi'\rangle$) with $N_\text{ex} + 1$ such excitations. 
Given that the reaction is thermally activated, 
we assume that the system is initially in 
a thermal distribution of reactant states, 
whose populations are denoted as $p_{(R,\chi)}^{(\text{th})}$.
For $N\leq 32$, $T$ = 298 K, and either with or without VSC, 
virtually all population
resides in $|R,0\rangle$ (Fig. \ref{fig:pop0}), 
the reactant state with 0 vibrational-cavity excitations.
Then the reaction kinetics should be governed purely by transitions from
$|R,0\rangle$ to product states with 
1 vibrational-cavity excitation [Fig. \ref{fig:rxn}(a)], 
as well as the corresponding backward transitions.

\begin{figure}[H]
\centering\includegraphics{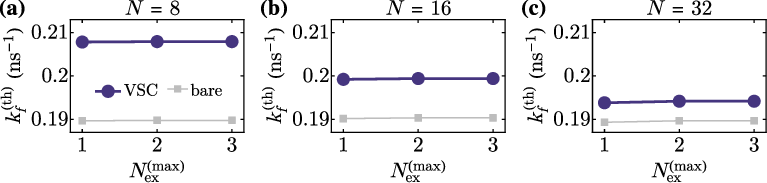}
	
\caption{\label{fig:kf}Rate ($k_f^{(\text{th})}$) of transition from 
a thermal distribution ($T$ = 298 K) of 
reactant states to all the product states 
as a function of $N_\text{ex}^\text{(max)}$
for VSC and bare reactions.
Rates are shown for (a) $N=8$, (b) $N=16$, and (c) $N=32$ 
and are calculated by averaging over 5000, 5000, and 256 disorder realizations, respectively.
Other calculation parameters are given in \S\ref{sec:default-parameters}.
}

\end{figure}

\begin{figure}[H]
\centering\includegraphics{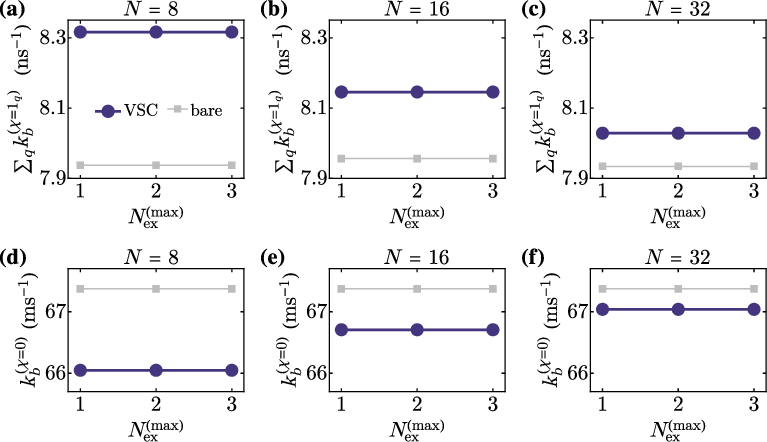}
	
\caption{\label{fig:kb}Total rate of transition to the reactant states from 
product states with either 1 [(a)-(c), total rate $\sum_q k_b^{(\chi=1_q)}$] or 
0 [(d)-(f), total rate $k_b^{(\chi=0)}$] vibrational-cavity excitations as 
a function of $N_\text{ex}^\text{(max)}$ for VSC and bare reactions.
Rates are shown for (a),(d) $N=8$, (b),(e) $N=16$, and (c),(f) $N=32$ 
and are calculated by averaging over 5000, 5000, and 256 disorder realizations, respectively.
Other calculation parameters are given in \S\ref{sec:default-parameters}.
}

\end{figure}

To numerically verify this statement for the VSC and bare reactions, 
we calculate
$k_f^{(\text{th})} = \sum_{\chi,\chi'} p_{(R,\chi)}^{(\text{th})} k_{(R,\chi)\rightarrow (P,\chi')}$,
the rate of transition from a thermal distribution of reactant states to all product states,
assuming states have at most $N_\text{ex}^{(\text{max})}$ vibrational-cavity excitations.
As $N_\text{ex}^{(\text{max})}$ is increased from 1 to 3, 
$k_f^{(\text{th})}$ does not change appreciably (Fig. \ref{fig:kf}).
A similar outcome is reached for the backward reactive transitions. 
Define $k_b^{(\chi)} = \sum_{\chi'} k_{(P,\chi)\rightarrow (R,\chi')}$
as the transition rate from $|P,\chi\rangle$ to 
all reactant states. 
Fig. \ref{fig:kb} plots
$\sum_q k_b^{(\chi=1_q)}$ [(a)-(c)] and $k_b^{(\chi=0)}$ [(d)-(f)],
which are the total rates of backward reactive transitions from 
product states with $N_\text{ex} = 1$ and $N_\text{ex} = 0$, respectively. 	
The rates computed with $N_\text{ex}^{(\text{max})} = 1$ are 
virtually identical to those computed with 
$N_\text{ex}^{(\text{max})} = 2,3$.
Given that the nonreactive transitions 
(\S\ref{sec:kinetic-model}) do not efficiently channel population 
from states with $N_\text{ex} = 0, 1$ to states with $N_\text{ex} \geq 2$,
we conclude that the multiply-excited states do not contribute significantly to the reaction 
and can therefore be neglected.

We emphasize that the above results are obtained 
using the parameters of \S\ref{sec:default-parameters}. 
The findings are expected to hold for the other 
parameter ranges explored in this work. 
We carry out calculations for $T < 298$ K, 
for which the reaction should be less affected by
the states with $N_\text{ex} \geq 2$ compared to the case of $T=298$ K.  
In addition, we vary
light-matter coupling, cavity detuning, and inhomogeneous broadening.
However,
energies and thermal occupations
should not change enough to warrant inclusion of the higher vibrational-cavity excitation manifolds.

\section{\label{sec:valid-largeN}Validity of the kinetic model for large $N$} 

In the main text, it is briefly discussed that our kinetic model (\S\ref{sec:kinetic-model})
is not valid when the number $N$ of molecules is sufficiently large. 
We now elaborate on this matter. 

At the core of our kinetic model is master equation (\ref{eq:master-eq}), 
which is an example of the Pauli master equation (PME) \citep{BreuerBook,MayBook}. 
The PME is a Markovian master equation that describes the exchange of 
population among the eigenstates of a quantum system, 
which interacts with an environment. 
Let $H_\text{tot}$ be the total Hamiltonian governing the system and environment.
Make the partition $H_\text{tot}=H_{0,S}+(H_\text{tot}-H_{0,S})$, where $H_{0,S}$ is 
a zeroth-order system Hamiltonian and $H_\text{tot}-H_{0,S}$ acts perturbatively on 
the $H_{0,S}$ eigenstates $\{|n\rangle\}$, which have frequencies $\{\omega_n\}$.
While $H_{0,S}$ is not necessarily 
the full Hamiltonian of the system in the absence of the bath, 
we assume that $H_\text{tot}-H_{0,S}$ gives rise to purely dissipative dynamics.
Then the PME can be written as
\begin{equation}
\frac{d\rho_{nn}(t)}{dt} = \sum_m k_{nn,mm} \rho_{mm}(t),
\label{eq:PME}
\end{equation}
where $\rho$ is the reduced density matrix of the system, 
$\rho_{nn}$ ($\rho_{mm}$) is the population of $H_{0,S}$ eigenstate $|n\rangle$ ($|m\rangle$),
and $k_{nn,mm}$ is the rate constant for population transfer from $|m\rangle$ to $|n\rangle$.
Notice that the populations are not coupled
to coherences, which are the off-diagonal elements $\rho_{lm}$, 
where $|l\rangle$ and $|m\rangle$ are different eigenstates of $H_{0,S}$.
In general, such coupling exists and cannot be ignored.

To understand when this decoupling is acceptable, 
consider the Redfield equation (RE) \citep{MayBook}, 
a Markovian quantum master equation that is more general than the PME.
For the proceeding analysis, it is convenient to first move to the interaction picture 
where the reduced density matrix evolves as
\begin{equation}
\rho^{(I)}_{lm}(t) = \langle l |e^{iH_{0,S}t}\rho(t)e^{-iH_{0,S}t}  |m\rangle = e^{i\omega_{lm}t}\rho_{lm}(t),
\label{eq:int-pic}
\end{equation}
where $\omega_{lm} = \omega_l - \omega_m$.
With this transformation, the RE, for populations of $H_{0,S}$ eigenstates, can be written as 
\begin{equation}
\frac{d\rho_{nn}(t)}{dt} = \sum_{m} k_{nn,mm} \rho_{mm}(t) + 
\sum_{l \neq m} k_{nn,lm} e^{-i\omega_{lm}t}\rho_{lm}^{(I)}(t), 
\label{eq:redfield}
\end{equation}
where $l, m$ run over the eigenstates of $H_{0,S}$.
In writing Eq. (\ref{eq:redfield}), we have used $\rho_{nn}^{(I)}(t) = \rho_{nn}(t)$ [Eq. (\ref{eq:int-pic})]. 
Compared to the PME [Eq. (\ref{eq:PME})], 
the RE [Eq. (\ref{eq:redfield})] additionally includes coupling of 
populations to coherences through the constants $k_{nn,lm}$.
To see when we can ignore this coupling, 
formally integrate Eq. (\ref{eq:redfield}):
\begin{equation}
\rho_{nn}(t) = \rho_{nn}(0) + \sum_{m} k_{nn,mm} \int_0^t dt' \, \rho_{mm}(t')  
+ \sum_{l \neq m} k_{nn,lm} \int_0^t dt' \, e^{-i\omega_{lm}t'}\rho_{lm}^{(I)}(t'). 
\end{equation}
Focus on the second summation, which runs over the coherences. 
If $e^{-i\omega_{lm}t'}$ oscillates much faster than the (relaxation) timescale $\tau_{\text{rel},lm}$  over 
which $\rho_{lm}^{(I)}(t')$ evolves,
then the integral in term $lm$ approximately vanishes for $t = O(\tau_{\text{rel},lm})$. 
Then a good approximation is to neglect all terms $lm$ that satisfy this condition, 
which can be concisely expressed as 
\begin{equation}
\frac{\omega_{lm} \tau_{\text{rel},lm}}{2\pi} \gg 1.
\label{eq:sa}
\end{equation} 
This approximation, which falls under
the secular approximation (also known as the rotating wave approximation) of
the theory of open quantum systems \citep{BreuerBook,MayBook},
can be heuristically interpreted as follows:
if the energy gap between two states is larger than 
their decay linewidths [Eq. (\ref{eq:sa})],
then the two states are ``good eigenstates" of the (reduced) system, 
and population will remain in these states in the presence of decoherence processes. 
Applying the secular approximation to 
all coherence terms, \emph{i.e.}, dropping all terms in the second summation of Eq. (\ref{eq:redfield}),
converts the RE to the PME.
In other words, the PME is valid when the secular approximation [Eq. (\ref{eq:sa})] holds for all coherences.
Thus, one should be careful when using the PME to model systems with (near) degeneracies.

In light of this, we reflect on the appropriateness of our kinetic model.
Here, a system of $N$ molecules features a quasidegenerate manifold of $N-1$ dark states, 
whose frequencies are approximately normally distributed with mean $\overline{\omega}_v$ and 
standard deviation $\sigma_v$ (see main text).
In our kinetic simulations, 
we vary $N$ while fixing $\sigma_v$.
As $N$ increases, the frequency spacing between dark states decreases.
At the same time, the rates of relaxation processes---which 
include the electron transfer reaction, vibrational/cavity decay, and 
scattering among polariton and dark states---do not change as much.
These relaxation rates determine the timescales over 
which coherences between dark states evolve [in the interaction picture of Eq. (\ref{eq:int-pic})].
So, the secular approximation, and therefore our kinetic model,
will be invalid for large enough $N$.

Given the default parameters used in our reaction-rate calculations
(\S\ref{sec:default-parameters}), let us estimate the values of $N$ for 
which the secular approximation holds. 
Motivated by the previous paragraph, we introduce a simplified version of 
the secular-approximation criterion [Eq. (\ref{eq:sa})]:
\begin{equation}
\frac{\Delta \omega_{\text{dark}} \tau_{\text{rel,\,dark}}}{2\pi} \gg  1,
\label{eq:sa_dark}
\end{equation}
where $\Delta \omega_{\text{dark}}$ is the characteristic frequency spacing between
consecutive dark states, and $\tau_{\text{rel,\,dark}}$ is the characteristic timescale for the evolution of
dark-state coherences [in the interaction picture of Eq. (\ref{eq:int-pic})].
Since the vast majority of dark states have frequency within $2\sigma_v$ of the mean ($\omega_v$),
we take $\Delta \omega_{\text{dark}} = 4\sigma_v/(N-1)$.
Accounting for only the aforementioned relaxation processes 
(\emph{i.e.}, ignoring pure dephasing \citep{MayBook}), 
we evaluate $\tau_{\text{rel,\,dark}}$ as the sum of 
characteristic rate constants for the population decay of a single dark state
(the corresponding quantity for population gain is smaller; see main text):
\begin{equation}
\tau_{\text{rel,\,dark}}^{-1} = \frac{k_b}{2} + \gamma +  k_\text{scat,\,dark}.
\end{equation}
The right-hand side is a sum of major decay rates,
where $k_b/2$ represents the rate of a backward reactive transition for
a dark state with 50\% character of the reactive mode,
$\gamma$ is the (bare) vibrational decay rate,
and $k_\text{scat,\,dark}$ is a characteristic rate at which a dark state decays into other dark states.
From the default parameters (\S\ref{sec:default-parameters}),  
we have $k_b \approx \gamma = 0.01$ ps$^{-1}$.
To estimate $k_\text{scat,\,dark}$,  
we consider three dark states ($q=1,2,3$) that are consecutive in frequency
and let $k_\text{scat,\,dark}$ be the sum of rates [Eq. (\ref{eq:k_redis})] of decay from dark state $q=2$ to 
each dark state $q' \neq 2$ (including states with $q'\neq 1,3$).
For simplicity and based on the main text, 
we assume that dark states $q=1,2,3$ are equally delocalized across 
two of four bare vibrational states ($i=1,2,3,4$),
such that the expansion coefficients $c_{qi}$ (see main text) satisfy $|c_{qi}|^2=0.5$ for
$(q,i)=(1,1),(1,2),(2,2),(2,3),(3,3),(3,4)$ [and $|c_{qi}|^2=0$ for all other $(q,i)$ where $q=1,2,3$].
Also, we set $\Delta \omega_{\text{dark}}$ as the frequency difference between 
states $q=1,2$ and between states $q=2,3$.
With $J(\omega)$ of Eq. (\ref{eq:j}) as the spectral density and
$T=298$ K as the temperature, 
Eq. (\ref{eq:k_redis}) yields
\begin{subequations}
\begin{align}
k_\text{scat,\,dark} &= 0.25 \mathcal{R}(-\Delta \omega_{\text{dark}} ) + 0.25 \mathcal{R}(\Delta \omega_{\text{dark}} ) \\
&\approx 0.5 \frac{0.05\text{ ps}^{-1}}{10\text{ cm}^{-1}}\Delta \omega_{\text{dark}}.
\end{align}
\end{subequations}
In obtaining the second line, we have used Fig. \ref{fig:r} to approximately linearize
$\mathcal{R}(\mp\omega)$ [Eq. (\ref{eq:r})] 
as $(0.05\text{ ps}^{-1})\omega/(10\text{ cm}^{-1})$ for $\omega \in [0,10]$ cm$^{-1}$.
Using the above rates, $\sigma_v = 10$ cm$^{-1}$ (\S\ref{sec:default-parameters}), and Eq. (\ref{eq:sa_dark}), 
we estimate that 
the secular approximation holds for $N \ll 74$.
In all numerical kinetic simulations, we set $N \leq 32$.

\section{Numerical kinetic simulations and rate calculations \label{sec:kinetic-sims_rate-calcs}}

In this section, we first describe our numerical simulations of the
reaction kinetics. We then discuss how we obtain the reaction rate
from the numerically determined reactant population versus time. To
be clear, we note that the quantities (e.g., populations, energies)
shown below pertain only to states with $\chi=0,1_{q}$ for $q=1,\dots,N+1$
(see main text).

State populations as a function of time are simulated by numerically
solving master equation (\ref{eq:master-eq}). We start by writing
the equation as $d\mathbf{p}/dt=A\mathbf{p}$, where $\mathbf{p}$
is the vector of populations and $A$ is the matrix of transition
rates. Subsequently, we apply the following standard procedure to
evaluate $\mathbf{p}(t)=\exp(At)\mathbf{p}(0)$ \citep{Frankcombe2001}.
This method employs the symmetrization of $A$ to avoid a numerically
unstable matrix inversion. First, we compute the matrix $B=MAM^{-1}$,
where $M$ is the diagonal matrix with diagonal elements $M_{(X,\chi),(X,\chi)}=f_{(X,\chi)}^{-1/2}$,
and $f_{(X,\chi)}=\exp(-\beta E_{(X,\chi)})/\sum_{(X,\chi)}\exp(-\beta E_{(X,\chi)})$.
Since the transition rates {[}Eqs. (\ref{eq:k_ET}), (\ref{eq:k_decay})-(\ref{eq:k_redis}){]}
satisfy detailed balance, $B$ is symmetric. After numerically diagonalizing
$B$, the population at time $t$ is evaluated as
\begin{equation}
\mathbf{p}(t)=M^{-1}Q\exp(Dt)Q^{\intercal}M\mathbf{p}(0),
\end{equation}
where $Q$ is a matrix whose columns are the eigenvectors of $B$,
$D$ is the diagonal matrix whose diagonal elements are the eigenvalues
corresponding to said eigenvectors, and $Q^{\intercal}=Q^{-1}$ due
to $B$ being symmetric. Because we are interested in thermally activated
reactivity, the vector of initial populations, $\mathbf{p}(0)$, is
taken to be a thermal distribution of reactant eigenstates:\begin{subequations}
\begin{align}
p_{(R,\chi)}(0) & =\frac{\exp(-\beta E_{(R,\chi)})}{\sum_{\chi}\exp(-\beta E_{(R,\chi)})},\\
p_{(P,\chi)}(0) & =0.
\end{align}
\label{eq:p0}
\end{subequations}We evaluate $\mathbf{p}(t)$ at $t=j\Delta t$,
where $\Delta t=0.2$ ns and $j=0,\dots,100$.

Next, the reaction rate is obtained by fitting the numerically determined
values of reactant population, 
\begin{equation}
p_{R}=\sum_{\chi}p_{(R,\chi)},
\end{equation}
and their respective values of $t$ to the exponential function 
\begin{equation}
p_{R}=\exp(-kt).
\end{equation}
The fitting parameter $k$ is the reaction rate. For all fits, the
adjusted $R^{2}$ values have mean 0.99999 and standard deviation
$7\times10^{-6}$. Such successful fitting reflects the reaction being
first-order \citep{AtkinsBook9th} in reactant. Here, first-order
kinetics occurs because product excited states do not accumulate sufficiently
(see $\S$\ref{sec:analytical_bare}) and the product ground state
does not revert to reactant states at a fast enough rate (due to high
activation energy). 

\section{Analytical rate: bare reaction \label{sec:analytical_bare}}

A simplified kinetic model for the bare reaction is shown in Fig.
\ref{fig:rxn}(a). In this model, the populations of $|R,0\rangle$
and $|P,1_{r}\rangle$ evolve as
\begin{align}
\frac{dp_{(R,0)}}{dt} & =-k_{f}p_{(R,0)}+k_{b}p_{(P,1_{r})},\label{eq:p_R0_bare}\\
\frac{dp_{(P,1_{r})}}{dt} & =-(k_{b}+\gamma)p_{(P,1_{r})}+k_{f}p_{(R,0)},\label{eq:p_P1r_bare}
\end{align}
respectively. Since $k_{f}\ll k_{b},\gamma$ (see main text), $p_{(P,1_{r})}$
does not accumulate, and so we apply the steady-state approximation
(SSA) \citep{AtkinsBook9th} to this population: $dp_{(P,1_{r})}/dt\approx0$.
Solving the resulting equation for $p_{(P,1_{r})}$ and plugging the
solution into Eq. (\ref{eq:p_R0_bare}) leads to $dp_{(R,0)}/dt\approx-k_{\text{bare}}^{(\text{analytical})}p_{(R,0)}$.
Defined in Eq. (\ref{eq:k_bare}), $k_{\text{bare}}^{(\text{analytical})}=k_{f}\left(\frac{\gamma}{\gamma+k_{b}}\right)$
represents the net rate of reactant depletion, \emph{i.e.}, the reaction
rate. 

\section{Analytical rate: VSC reaction\label{sec:analytical_VSC}}

Here, we consider a simplified kinetic model for the VSC reaction.
In this model, polaritons are decoupled from the reaction, and the
reaction proceeds through multiple reaction channels, each involving
the (de)excitation of a dark mode. Define $\mathcal{D}$ as the set
of dark modes. The populations of $|R,0\rangle$ and $|P,1_{q}\rangle$,
where $q\in\mathcal{D}$, evolve according to
\begin{align}
\frac{dp_{(R,0)}}{dt} & =-\sum_{q\in\mathcal{D}}k_{f}^{(q)}p_{(R,0)}+\sum_{q\in\mathcal{D}}k_{b}^{(q)}p_{(P,1_{q})},\label{eq:p_R0_VSC}\\
\frac{dp_{(P,1_{q})}}{dt} & =-\left(k_{b}^{(q)}+\gamma\right)p_{(P,1_{q})}+k_{f}^{(q)}p_{(R,0)},\label{eq:p_P1q_VSC}
\end{align}
respectively. In analogy to the derivation of $\S$\ref{sec:analytical_bare},
we can apply the SSA to each $p_{(P,1_{q})}$ to arrive at
\begin{equation}
\frac{dp_{(R,0)}}{dt}\approx-k_{f}\left(\sum_{q\in\mathcal{D}}|c_{qr}|^{2}\frac{\gamma}{\gamma+|c_{qr}|^{2}k_{b}}\right)p_{(R,0)},\label{eq:p_R0_VSC_2}
\end{equation}
where we have used Eq. (\ref{eq:k_f/b}). Eq (\ref{eq:p_R0_VSC_2})
is equivalent to $dp_{(R,0)}/dt\approx-k_{\text{VSC}}^{(\text{analytical})}p_{(R,0)}$,
where the reaction rate constant $k_{\text{VSC}}^{(\text{analytical})}=k_{f}\left(\sum_{q\in\mathcal{D}}|c_{qr}|^{2}\frac{\gamma}{\gamma+|c_{qr}|^{2}k_{b}}\right)$
is equivalent to that shown in Eq. (\ref{eq:k_VSC}) the main text.

\section{Thermodynamic parameters of activation}

To determine the thermodynamic parameters of activation for a given
reaction (\emph{i.e.}, set of reaction parameters excluding temperature,
$T$), we calculate the reaction rate, $k$, for $T=$ 278, 283, 288,
293, 298 K. The rates are computed using numerical kinetic simulations,
as described in $\S$\ref{sec:kinetic-sims_rate-calcs}. We then fit
the $(k,T)$ values to the Eyring-Polanyi equation \citep{Eyring1935jcp,Evans1935},
\begin{equation}
k=\frac{k_{B}T}{h}\exp\left(-\frac{\Delta H^{\ddagger}}{RT}+\frac{\Delta S^{\ddagger}}{R}\right),\label{eq:eyring-polanyi}
\end{equation}
where $R$ is the gas constant. The fitting parameters $\Delta H^{\ddagger}$
and $\Delta S^{\ddagger}$ are the enthalpy and entropy, respectively,
of activation.

For all fits, the adjusted $R^{2}$ values have mean 0.9999996 and
standard deviation $1\times10^{-7}$. This excellent agreement between
the numerically obtained rates and the Eyring-Polanyi equation is
attributed to a fortuitous choice of temperature range over which
the fittings are performed. Suboptimal goodness of fit is expected
for general ranges of $T$ because the transition rates {[}Eqs. (\ref{eq:k_ET}),
(\ref{eq:k_decay})-(\ref{eq:k_redis}){]} of our kinetic model have
a different functional form (with respect to $T$) compared to Eq.
(\ref{eq:eyring-polanyi}).

\section{effect of inhomogeneous broadening on the reaction under VSC}

We carry out reaction simulations for two
additional values of inhomogeneous broadening, $\sigma_v = 15, 20$ cm$^{-1}$. 
These $\sigma_v$ are slightly higher than that used in 
the simulations of the main text, \emph{i.e.}, $\sigma_v=10$ cm$^{-1}$.

\begin{figure}[h]
	\centering\includegraphics{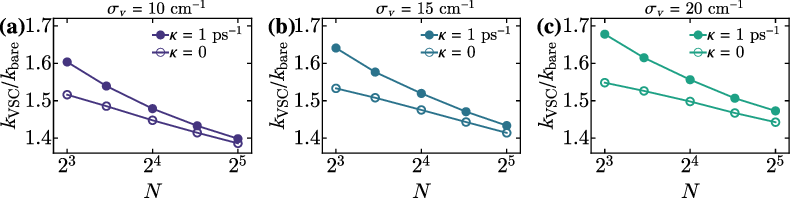}
	
	\caption{\label{fig:rxn_sigma-nS}$k_{\text{VSC}}/k_{\text{bare}}$
			as a function of $N$ for fixed light-matter coupling strength $g\sqrt{N}=80$ cm$^{-1}$, 
			various rates $\kappa$ of cavity decay, and
			inhomogeneous broadening (a) $\sigma_v=10$ cm$^{-1}$, (b) $\sigma_v=15$ cm$^{-1}$, 
			and (c) $\sigma_v=20$ cm$^{-1}$.
			The rates $k_{\text{VSC}}$ and $k_{\text{bare}}$ are averages over 5000 disorder realizations.}
	
\end{figure}

\begin{figure}[h]
	\centering\includegraphics{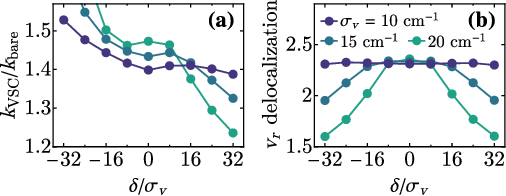}
	
	\caption{\label{fig:ratio-deloc_sigma-detuning}(a) $k_{\text{VSC}}/k_{\text{bare}}$
			and (b) $v_{r}$ delocalization, as a function of cavity detuning $\delta$, 
			for $N = 2^5$, fixed collective light-matter coupling strength $g\sqrt{N}=80$ cm$^{-1}$, 
			and various values of inhomogeneous broadening $\sigma_v$. 
			In (a), $k_{\text{VSC}}$ and $k_{\text{bare}}$ are averages over 5000 disorder realizations.}
	
\end{figure}

\begin{figure}[h]
\centering\includegraphics{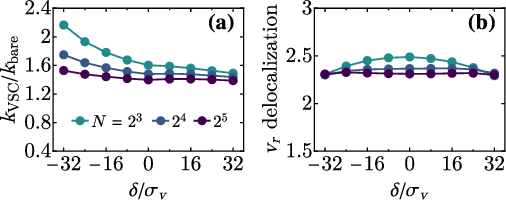}
	
\caption{\label{fig:ratio-deloc_nS-g80}(a) $k_{\text{VSC}}/k_{\text{bare}}$ and (b) $v_{r}$ delocalization,
as a function of cavity detuning $\delta$, for various $N$ and fixed
collective light-matter coupling strength $g\sqrt{N}=8\sigma_{v}$.
In (a), $k_{\text{VSC}}$ and $k_{\text{bare}}$ are averages 
over 5000 disorder realizations.
For $\delta \ll 0$, $k_{\text{VSC}}/k_{\text{bare}}$ has a significant positive contribution from
the reduced activation energy afforded by the lower polariton \citep{Campos-Gonzalez-Angulo2019}, 
whose overlap with $v_{r}$ scales as $O(N^{-1})$.
As $N$ increases, this polaritonic contribution diminishes, 
and the $\delta$-dependence of $k_{\text{VSC}}/k_{\text{bare}}$
becomes more similar to that of $v_{r}$ delocalization.
}
	
\end{figure}

For each of the three $\sigma_v$ values,
\begin{enumerate}
\item the effect of cavity decay on the reaction is minor and diminishes with $N$ 
(Fig. \ref{fig:rxn_sigma-nS}).
\item the reaction-rate enhancement [Fig. \ref{fig:ratio-deloc_sigma-detuning}(a)] follows the same behavior 
as the reactive-mode delocalization [Fig. \ref{fig:ratio-deloc_sigma-detuning}(b)] 
when cavity detuning is varied  
(except at large negative detuning, when the lower polariton significantly affects the reaction;
see Fig. \ref{fig:ratio-deloc_nS-g80}),
\end{enumerate}
Hence, the positive correlation between reactivity under VSC and 
semilocalization of the dark modes (see main text) 
is also unchanged when inhomogeneous broadening is moderately increased from the value in the main text. 
This statement is further supported by the following:
increasing inhomogeneous broadening [Fig. \ref{fig:ratio-deloc_sigma-detuning}(a)] and 
reducing light-matter coupling [Fig. \ref{fig:ratio-deloc_g-detuning}(a)] affect the reaction rate similarly
(except at negative detuning; see item 2 above).
Indeed, the molecular PR of the dark modes depends on $g\sqrt{N}/\sigma_v$.

\section{additional supplemental figures}

\bibliographystyle{apsrev4-2}

\makeatletter\@input{auxSI.tex}\makeatother